\documentclass[
  a4paper,
  aps,pre,
  superscriptaddress,
  floatfix,
  nofootinbib,
  longbibliography,
  notitlepage,
  twocolumn
]{revtex4-1}

\usepackage{amsmath,amssymb,amsfonts}
\usepackage{mathtools}
\usepackage{bm}
\usepackage{color}
\usepackage{latexsym}
\usepackage{verbatim}
\usepackage{multirow}
\usepackage{rotating}
\usepackage{lipsum}
\usepackage[english]{babel}
\usepackage{comment}
\usepackage{braket}
\usepackage{blkarray,bigstrut}
\usepackage{bbold}  
\usepackage{hyperref}

\usepackage{url}
\usepackage{hyperref}

\newcommand{\DEL}[1]{{{\color{cyan}#1}}}
\renewcommand{\DEL}[1]{}

\begin{document}

\title{Emergence of higher-order interactions in systems of coupled Kuramoto oscillators with time delay}

\author{Narumi Fujii}
\email{Corresponding author: fujii.n.801e@m.isct.ac.jp}
\affiliation{Department of Systems and Control Engineering, Institute of Science Tokyo (former Tokyo Tech), Tokyo 152-8552, Japan}
\affiliation{Department of Mathematics \& naXys, Namur Institute for Complex Systems, University of Namur, Namur B5000, Belgium}

\author{Keisuke Taga}
\affiliation{Department of Systems and Control Engineering, Institute of Science Tokyo (former Tokyo Tech), Tokyo 152-8552, Japan}
\affiliation{Department of Physics and Astronomy, Tokyo University of Science, Chiba 278-8510, Japan}

\author{Riccardo Muolo}
\affiliation{Department of Systems and Control Engineering, Institute of Science Tokyo (former Tokyo Tech), Tokyo 152-8552, Japan}
\affiliation{RIKEN Center for Interdisciplinary Theoretical and Mathematical Sciences (iTHEMS), Saitama 351-0198, Japan}

\author{Bob Rink}
\affiliation{Department of Mathematics, Vrije Universiteit Amsterdam, Amsterdam 1081 HZ, The Netherlands}

\author{Hiroya Nakao}
\affiliation{Department of Systems and Control Engineering, Institute of Science Tokyo (former Tokyo Tech), Tokyo 152-8552, Japan}
\affiliation{International Research Frontiers Initiative, Institute of Science Tokyo (former Tokyo Tech), Kanagawa 226-8501, Japan}

\date{\today}

\begin{abstract}
We show that higher-order interactions naturally emerge from time-delayed pairwise coupling in Kuramoto oscillators.
By expanding the delayed pairwise coupling to the second order, we derive a delay-free Kuramoto model possessing both pairwise and three-body interactions. Numerical simulations and stability analysis demonstrate that
the three-body Kuramoto model and the time-delayed pairwise Kuramoto model exhibit qualitatively consistent synchronization transitions under appropriate conditions.
In particular, the bistability arising in the time-delayed Kuramoto model is accounted for by the three-body interactions.
Our findings reveal that time delays can be recast effectively as higher-order interactions, providing an insight into how coupling delays shape collective dynamics.
\end{abstract}

\maketitle

\section{Introduction}

Increasing evidence suggests that interactions in real-world systems are inherently higher-order (many-body), and can be naturally modeled through hypergraphs and simplicial complexes as extensions of traditional networks~\cite{battiston2020networks,bianconi2021higher,natphys,Dibakar_2022,bick2023higher,boccaletti2023structure,muolo2024turing,millan2025topology,battiston2025collective}.
These higher-order interactions deeply affect the dynamics of complex systems, enriching their behavior. Such effects have been observed in a wide range of dynamical processes, such as synchronization~\cite{millan2020explosive,gambuzza2021stability,gallo2022synchronization,von2023hypernetworks,von2024higher,zhang2024deeper}, swarming~\cite{anwar2024collective,hu2025effect,leon2025collective}, random walks~\cite{carletti2020random,schaub2020random}, stochastic dynamics~\cite{wang2025network}, pattern formation~\cite{Muolo2023turing,carletti2020dynamical}, opinion dynamics~\cite{iacopini2019simplicial,Neuhauser2020multibody}, and control~\cite{de2022pinning,della2023emergence,xia2024pinning}, to name a few.
\par

For instance, the classical
Kuramoto model~\cite{Kuramoto1975self} with pairwise coupling typically exhibits a transition from incoherence to 
synchrony as the 
coupling strength increases~\cite{Kuramoto1984chemical,acebron2005kuramoto}.
Introducing higher-order coupling into the Kuramoto model leads to
richer dynamics, such as hysteresis, bistability between incoherence and synchrony, cluster formation, and slow switching~\cite{tanaka2011multistable,Skardal2020higher,lucas2020multiorder,Leon2024higher,Leon2025theory}.
Kuramoto models with higher-order interactions are obtained by 
phase reduction~\cite{Nakao2016phase,Monga2019phase1,pietras2019network} of limit-cycle oscillators with  many-body interactions~\cite{Leon2024higher,leon2025collective}. They also arise from second-order phase reduction of limit-cycle oscillators with purely pairwise interactions~\cite{Kuramoto_2019,Leon2019phase,gengel2020high,nijholt2022emergent,mau2023high,mau2024phase,bick2024higher}.
Parametric phase reduction technique~\cite{von2023parametrisation} for delayed Kuramoto-type models has also been developed recently~\cite{bick2024time,bick2025higher}.
\par
In this study, we reveal that time-delayed interactions, which are ubiquitous in real-world system, can also be interpreted as effective higher-order interactions. 
Starting from a Kuramoto model with time-delayed pairwise interactions~\cite{Yeung_1999}, used as models of coupled neurons~\cite{ermentrout2007delay}, lasers~\cite{Kozyreff2000Lasers}, slime molds~\cite{Takamatsu2000Living}, and power grids~\cite{Taher2019grid,Bottcher2020grid}, 
we demonstrate that the structure of higher-order interactions naturally emerges from delayed interactions.
Furthermore, by numerical simulations and analysis via the Ott–Antonsen (OA) ansatz \cite{Ott_Antonsen_2008}, we show that the bistability between synchronized and incoherent states arising in the time-delayed Kuramoto model is accounted for by effective three-body interactions.
\par

We emphasize that our aim is not to provide an accurate approximation scheme for the time-delayed Kuramoto model by higher-order expansions, but rather to juxtapose the higher-order (three-body) interactions and time-delayed pairwise interactions.
We reveal that these interactions, which may be of distinct physical origins, lead to qualitatively similar collective dynamics.

This paper is organized as follows.
In Sec. II, we derive the higher-order Kuramoto model from the time-delayed Kuramoto model and compare their dynamics numerically in Sec. III. In Sec. IV, we analyze their stability via the OA ansatz and compare their phase diagrams. Sec. V concludes the paper and the Appendices provide detailed calculations.

\section{Emergence of Higher-Order Interactions from Time Delay} 

We consider a system of globally coupled phase oscillators with time delay,
given by the following set of delay-differential equations:
\begin{align}
    \dot{\theta}_j(t)=\omega_j+\frac{\epsilon}{N}\sum_{\substack{k=1\\k\neq j}}^{N}\sin\left(\theta_k(t-\tau)-\theta_j(t)\right),
    \label{eq:Kuramoto_time_delay}
\end{align}
for $j=1, ..., N$, where $N$ is the number of the oscillators, $\theta_j$ and $\omega_j$ are the phase and natural frequency of $j$th oscillator, respectively, and
$\omega_j$ follows a unimodal probability density function $g(\omega)$
with central frequency $\omega_0$.
In the coupling term, $\epsilon \geq 0$ is the coupling strength, 
and $\tau \geq 0$ is a time delay in the transmission of the phase information. 
We assume that no self coupling exists and the delay 
is uniform over all oscillator pairs.
This model was analyzed in detail by Yeung and Strogatz in~\cite{Yeung_1999}.
We assume that the coupling strength $\epsilon$ and the time delay $\tau$ are moderately small.

We begin by rewriting Eq.~\eqref{eq:Kuramoto_time_delay} as
\begin{equation}
    \dot{\theta}_j(t)
    = \omega_j+\frac{\epsilon}{N}\sum_{\substack{k=1\\k\neq j}}^{N}
    \sin\left(\theta_k(t) - \int_0^{\tau} \dot \theta_k(t-s)ds -\theta_j(t)\right).
    \label{eq:taylor_expansionchanged}
\end{equation}
Substituting the perturbative approximation of $\dot \theta_k(t-s)$ into Eq.~\eqref{eq:taylor_expansionchanged}, expanding the sine function, and evaluating the time integrals, we obtain the following set of ordinary differential equations approximating Eq.~\eqref{eq:Kuramoto_time_delay} up to $\mathcal{O}(\epsilon^2)$ (see Appendix \ref{subsec:derivation_higher-order} for details): 
{\small
\begin{widetext}
\begin{align}
    \dot{\theta}_j(t) =&\ \omega_j+\frac{\epsilon}{N}\sum_{\substack{k=1\\k\neq j}}^{N}\sin\left(\theta_k(t)-\theta_j(t)
    -\omega_k\tau\right)\notag\\
    &+\frac{\epsilon^2}{N^2}\sum_{\substack{k=1\\k\neq j}}^{N}\sum_{\substack{l=1\\l\neq k}}^{N}
    \cos\left(\theta_k(t)-\theta_j(t)-\omega_k\tau\right)
    \Big\{-\sin\left(\theta_l(t)-\theta_k(t)-\omega_l\tau\right)
    f_1(\omega_l, \omega_k)+\cos\left(\theta_l(t)-\theta_k(t)-\omega_l\tau\right)
    f_2(\omega_l, \omega_k)\Big\}+\mathcal{O}(\epsilon^3), 
    \label{eq:dynamics_second}
\end{align}
\end{widetext}}%
\noindent where the functions $f_1$ and $f_2$ are defined as
\begin{align}
f_1(\omega_l, \omega_k)&=
\frac{\sin\left((\omega_l-\omega_k)\tau\right)}{{\omega_l-\omega_k}}
\; (\omega_l \neq \omega_k),
\quad
\tau
\; (\omega_l=\omega_k),
\end{align}
and
\begin{align}
f_2(\omega_l, \omega_k)&=
\frac{1-\cos\left((\omega_l-\omega_k)\tau\right)}{\omega_l-\omega_k}
\; (\omega_l \neq \omega_k),
\quad
0
\; (\omega_l=\omega_k),
\end{align}
respectively.
This represents a \textit{higher-order Kuramoto model} with pairwise and three-body interactions.

\color{black}

On the right-hand side of Eq.~\eqref{eq:dynamics_second}, the $\mathcal{O}(\epsilon)$ term represents pairwise interactions, while the $\mathcal{O}(\epsilon^2)$ term corresponds to three-body interactions, each with phase lags.
Higher-order interactions involving four or more bodies can also arise at $\mathcal{O}(\epsilon^3)$ and beyond, which are 
neglected in what follows. 
In the case of identical oscillator frequencies, we obtain
\begin{widetext}
\begin{equation}
    \dot{\theta}_j(t)
    =\omega_0+\frac{\epsilon}{N}\sum_{\substack{k=1\\k\neq j}}^{N}\sin\left(\theta_k(t)-\theta_j(t)
    -\omega_0\tau\right)+\frac{\epsilon^2\tau}{2N^2}\sum_{\substack{k=1\\k\neq j}}^{N}\sum_{\substack{l=1\\l\neq k}}^{N}
    \Big\{-\sin\left(\theta_l(t)-\theta_j(t)-2\omega_0\tau
    \right)+\sin\left(2\theta_k(t)-\theta_l(t)-\theta_j(t)\right)\Big\}.
    \label{eq:dynamics_second_identical}
\end{equation}
\end{widetext}

In this model, the pairwise interaction is 
of the Kuramoto-Sakaguchi type~\cite{Sakaguchi_1986, ermentrout1997delay} with a phase lag $\omega_0 \tau$, while the three-body interaction is given by the $(2,-1,-1)$ harmonic~\cite{Skardal2020higher,namura2025optimal}.
The higher-order interaction can be interpreted as the influence of oscillator $l$ on oscillator $j$ mediated through oscillator $k$. 
This effect is incorporated by modifying the Sakaguchi--Kuramoto interaction as 
$\sin\left(\theta_k - \theta_j - \omega_0 \tau + \delta_{kl}\right)
\simeq \sin\left(\theta_k - \theta_j - \omega_0 \tau\right)
+ \delta_{kl} \cos\left(\theta_k - \theta_j - \omega_0 \tau\right)$, 
where $\delta_{kl}\simeq \epsilon\sin\left(\theta_l - \theta_k - \omega_0 \tau\right)$ represents the influence of oscillator $l$ on oscillator $k$. 
The cosine term then represents the sensitivity to the perturbation $\delta_{kl}$, 
and thus leads to the structure of higher-order interactions in Eq.~\eqref{eq:dynamics_second_identical}.
Note that $\epsilon^2\tau$ appears as the strength of
the three-body interaction in Eq.~\eqref{eq:dynamics_second_identical}.
The order of the next term is $\epsilon^3 \tau^2$, hence the accuracy of this approximation is characterized by the smallness of $\epsilon \tau$.
Our aim in this study is not to pursue higher-order expansions further, but rather to compare the delay-free three-body model obtained above with the time-delayed pairwise model to reveal that they lead to qualitatively similar collective dynamics.
\par

\begin{figure}[tb]
    \centering
    \includegraphics[width=0.8\linewidth]{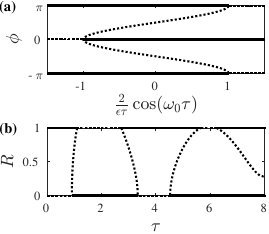}
    \caption{Synchronization transitions of $N=2$ oscillators for $\omega_0=\pi/2$ and $\epsilon=0.3$.
    (a) Fixed points of $\phi$ vs.
    $\frac{2}{\epsilon\tau}\cos(\omega_0 \tau)$.
    (b) Order parameter $R$ vs. $\tau$.
    The solid and dotted lines represent the stable and unstable branches, respectively.}
    \label{fig:transitions_for2}
\end{figure}

Before proceeding to the many-oscillator case, we show that the higher-order term 
can induce nontrivial synchronization transitions even for
two oscillators~\cite{Kori_2003}.
From Eq.~\eqref{eq:dynamics_second_identical} for $N=2$,
the phase difference $\phi=\theta_1-\theta_2$ obeys
\begin{equation}
    \dot{\phi}=-\frac{\epsilon^2\tau}{2}\sin\phi\left\{\frac{2}{\epsilon\tau}\cos\left(\omega_0\tau\right)+
    \cos\phi\right\},
    \label{eq:2_oscillators}
\end{equation}
which has in-phase ($\phi=0$) and anti-phase ($\phi=\pm\pi$) fixed points.
These fixed points are bistable in the regime $|\frac{2}{\epsilon\tau}\cos(\omega_0 \tau)|<1$ as shown in Fig.~\ref{fig:transitions_for2}(a).
Figure~\ref{fig:transitions_for2}(b) shows the same result,
where the order parameter $R$ (see below for the definition) is plotted as a function of $\tau$. 
Here, $R=0$ and $R=1$ correspond to the anti-phase (incoherent) and in-phase (fully synchronized) states, respectively. 
These fixed points exchange stability aperiodically as $\tau$ is increased, with bistable regimes arising near the transition points.
Similar behavior is observed also for the case with large $N$, as we show below.
For simplicity, we consider the case of identical oscillators in the following numerical simulations and analysis.

\begin{figure*}[!t]
    \centering
    \includegraphics[width=0.88\textwidth]{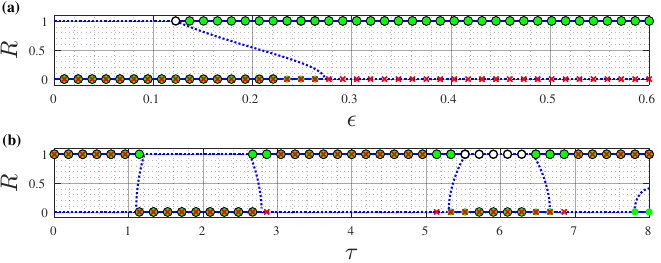}
    \caption{
    Synchronization transitions of the delayed Kuramoto model, Eq.~\eqref{eq:Kuramoto_time_delay} (black circles),
    higher-order Kuramoto model, Eq.~\eqref{eq:dynamics_second_identical} (green solid circles),
    and pairwise Kuramoto model (red crosses)
    obtained by numerical simulations for $N=300$ identical oscillators with $\omega_0=\pi/2$.
    (a) Order parameter $R$ vs. coupling strength $\epsilon$ for a fixed time delay $\tau=2.8$.
    (b) Order parameter $R$ vs. time delay $\tau$ for a fixed coupling strength $\epsilon=0.3$.
    The blue solid and dotted lines represent the stable and unstable branches of the 
    order parameter obtained from Eq.~\eqref{eq:dynamics_z_identical}.
    }
    \label{fig:transitions}
\end{figure*}

\section{Numerical Simulations }
Let us now assume $N \gg 1$ and
perform numerical simulations of the delayed Kuramoto model, Eq.~\eqref{eq:Kuramoto_time_delay}, and the higher-order Kuramoto 
model, Eq.~\eqref{eq:dynamics_second_identical}, in the case of identical oscillators. 
For comparison, we also show the results 
for the pairwise Kuramoto model without time delay, which 
neglects the higher-order term of $\mathcal{O}(\epsilon^2)$ in Eq.~\eqref{eq:dynamics_second_identical}.
To illustrate the collective behavior, we use the Kuramoto order parameter $z=R e^{i\Psi}=1/N\sum_{j=1}^N e^{i\theta_j}$,
where $0 \leq R \leq 1$ represents the degree of synchronization and $\Psi$ denotes the collective phase. 
\par
Figure~\ref{fig:transitions} shows the synchronization transitions
of the delayed, higher-order, and pairwise Kuramoto models
when the parameter $\epsilon$ or $\tau$ is varied, obtained by numerical simulations for $N=300$ (circles and crosses).
In Fig.~\ref{fig:transitions}(a), the order parameter $R$ is plotted as a function of the coupling strength $\epsilon$ 
at a fixed time delay $\tau=2.8$ for each model.
The delayed (black circles) and higher-order (green solid circles) Kuramoto models exhibit synchronization transitions from incoherence ($R=0$) to full synchrony ($R=1$) as $\epsilon$ is increased, while the pairwise Kuramoto model (red crosses) does not synchronize because the pairwise coupling is repulsive, $\pi/2 < \omega_0 \tau < 3\pi/2$.
In Fig.~\ref{fig:transitions}(b), $R$ is plotted against time delay $\tau$ at a fixed coupling strength $\epsilon = 0.3$.
All three models exhibit switching between full synchrony and incoherence as $\tau$ is increased.
\par

Note that the bistability between coherence and incoherence is observed only in the time-delayed and higher-order Kuramoto models, and not in the pairwise Kuramoto model.
Thus, three-body interactions are necessary to qualitatively account for the bistability
observed in the time-delayed Kuramoto model.
The fact that the three-body Kuramoto model reproduces this bistability suggests that the primary effect of time delay in pairwise coupling can be captured
as an effective equal-time three-body interaction.
We also note that slight quantitative discrepancies exist in between the higher-order and time-delayed models. This is expected as we truncated the expansion at the second order.

\section{Analysis via Ott-Antonsen ansatz}
For the higher-order Kuramoto model, Eq.~\eqref{eq:dynamics_second_identical},
we can derive macroscopic dynamics of the complex order parameter
to analyze the synchronization transitions.
We begin with Eq.~\eqref{eq:dynamics_second} as the microscopic model and
assume that each $\omega_j$ is independently chosen from an identical Lorentzian distribution given by $g(\omega)=(\gamma / \pi)\left[\left(\omega-\omega_0\right)^2+\gamma^2\right]^{-1}$.
Here, we assume 
$\gamma= \mathcal{O}(\epsilon) $ to approximate the case with identical oscillator frequencies.
We introduce general complex order parameters as
\begin{align}
    z_m=\frac{1}{N}\sum_{j=1}^{N}e^{i m (\theta_j - \omega_j\tau)}
    \label{eq:order_parameter_with_lag}
\end{align}
for $m=1, 2, ...$, where we include the phase lag $\omega_j \tau$ in the definition for convenience.
Then, Eq.~\eqref{eq:dynamics_second} is approximated as
\begin{align}
    \dot{\theta}_j(t)
    \simeq
    \omega_j+\frac{1}{2i}\left\{e^{-i\theta_j(t)}H_j(t)-e^{i\theta_j(t)}H_j^*(t)\right\},
    \label{eq:dynamics_second_order}
\end{align}
where
${}^*$ denotes the complex conjugate and $H_j=\epsilon z_1-\frac{\epsilon^2\tau}{2}\left(e^{-i\omega_j\tau}z_1-e^{i\omega_j\tau}z^*_{1}z_2\right)$ (see Appendix \ref{subsec:Eq8} for details).
\begin{figure*}[tb]
    \centering    
    \includegraphics[width=0.775\textwidth]{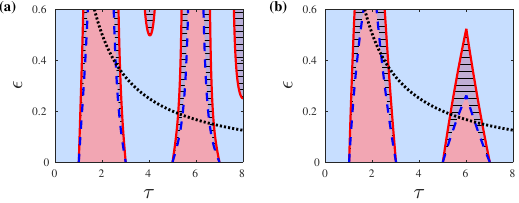}
    \caption{Stability diagrams of the fully synchronized state (blue regions) and incoherent state (red regions) for (a) the higher-order Kuramoto model, Eq.~\eqref{eq:dynamics_second_identical} via the OA ansatz,
    and for (b) the delayed Kuramoto model, Eq.~\eqref{eq:Kuramoto_time_delay}, obtained in~\cite{Yeung_1999}, for identical oscillators with $\omega_0=\pi/2$.
    The red solid lines and the blue dashed lines indicate the boundaries of these regions, and the black dotted line indicates $\epsilon\tau=1$.
    The purple stripes indicate the bistable regions of full synchrony and incoherence.}
    \label{fig:stability_zoom}
\end{figure*}
From this equation, we can derive the dynamics of the order parameter $z = z_1e^{i\omega_0\tau}$ in the $N \to \infty$ limit via the OA ansatz, and taking the $\gamma \to +0$ limit 
to approximate a population of homogeneous oscillators with identical frequencies, we obtain (see Appendix \ref{subsec:OA} for details)
\begin{widetext}
\begin{align}
    \dot{z}(t)
    &=\left\{i\omega_0+ \frac{\epsilon}{2}e^{-i\omega_0\tau}-
    \frac{\epsilon^2\tau}{4}
    e^{-2i\omega_0\tau}
    -\left(\frac{\epsilon}{2}e^{i\omega_0\tau}-\frac{\epsilon^2\tau}{4}-\frac{\epsilon^2\tau}{4}e^{2i\omega_0\tau}
    \right)|{z}(t)|^2-\frac{\epsilon^2\tau}{4}
    |{z}(t)|^4\right\}{z}(t).
    \label{eq:dynamics_z_identical}
\end{align}
\end{widetext}

Equation~\eqref{eq:dynamics_z_identical} for the complex order parameter has a fixed point at 
$z=0$, corresponding to the incoherent state of the oscillators,
and a limit cycle $z(t)=e^{i\Omega_s t}$
with $\Omega_s=\omega_0-\epsilon\sin\left(\omega_0\tau\right)+
\frac{\epsilon^2\tau}{2}\sin\left(2\omega_0\tau\right)$,
representing a fully synchronized state of the oscillators,
at any parameter values.
In addition, an intermediate limit cycle with a smaller amplitude,
$z(t)=\tilde{R} e^{i \Omega_i t}$
with $\tilde{R}=\sqrt{-\frac{2}{\epsilon\tau}\cos(\omega_0 \tau)+\cos(2\omega_0 \tau)}$ and $\Omega_i=\omega_0 -\left(\frac{\epsilon}{2}\sin(\omega_0 \tau)-\frac{\epsilon^2\tau}{4}\sin(2\omega_0\tau)\right)\left(1+{\tilde{R}^2}\right)$,
exists when $0\leq -\frac{2}{\epsilon\tau}\cos(\omega_0 \tau)+\cos(2\omega_0 \tau)\leq 1$.
Linear stability analysis of these solutions 
shows that the fixed point $z=0$ is stable when
\begin{align}
    \frac{\epsilon\tau}{2}\cos(2\omega_0 \tau)>\cos(\omega_0 \tau),
    \label{eq:stable_R0}
\end{align}
and the limit cycle $z(t)=e^{i \Omega_s t}$ is stable when
\begin{align}    
    \frac{\epsilon\tau}{2}\left(\cos(2\omega_0 \tau)-1\right)<\cos(\omega_0 \tau).
    \label{eq:stable_R1}
\end{align}
The intermediate limit cycle $z(t) = \tilde{R} e^{i \Omega_i t}$ is always unstable.
See Appendix \ref{subsec:linear_stability_OA} for details.
\par

The solid and blue dotted lines in Fig.~\ref{fig:transitions} show the stable and unstable solutions of Eq.~\eqref{eq:dynamics_z_identical}, showing good agreement with the results of numerical simulations for finite $N$.
Figure~\ref{fig:stability_zoom}(a) shows the stability regions described by Eqs.~\eqref{eq:stable_R0} and \eqref{eq:stable_R1}.
The incoherent state and fully synchronized state are bistable in the range where 
Eq.~(\ref{eq:stable_R0}) and (\ref{eq:stable_R1}) are simultaneously satisfied.
When the red solid line is crossed, a subcritical Hopf bifurcation occurs. At this point, the amplitude of the unstable limit cycle shrinks to zero, and the stability of the fixed point $z=0$ changes.
On the other hand, when the blue dashed line is crossed, a transcritical bifurcation occurs, and the stability of the limit cycle $z(t)=e^{i\Omega_s t}$ changes.
Thus, Eq.~\eqref{eq:dynamics_z_identical} 
analytically explains the synchronization transitions in Fig.~\ref{fig:transitions} for the higher-order Kuramoto model.

Figure~\ref{fig:stability_zoom}(b) shows the stability diagram of the incoherent and fully synchronized states for the delayed Kuramoto model in the $N \to \infty$ limit obtained by Yeung and Strogatz~\cite{Yeung_1999} for comparison.
The stability diagram of the higher-order Kuramoto model in Fig. \ref{fig:stability_zoom}(a) and that of the delayed Kuramoto model in Fig. \ref{fig:stability_zoom}(b) agree well
when $\epsilon\tau$ is small, while they exhibit
considerable discrepancy when $\epsilon\tau>1$.
Thus, the delayed Kuramoto model in Eq.~\eqref{eq:Kuramoto_time_delay} 
with identical frequencies
and the higher-order Kuramoto model in Eq.~\eqref{eq:dynamics_second_identical} undergo qualitatively consistent synchronization transitions when $\epsilon\tau$ is small.

We note that the time-delayed Kuramoto model has been analyzed in pioneering works~\cite{Ott_Antonsen_2008, Ott_Antonsen_2009} by directly applying the OA ansatz.
Equation~\eqref{eq:dynamics_z_identical} for the complex order parameter can also be obtained by expanding the time-delayed term in the delay differential equation for the macroscopic order parameter using their results; see Appendix \ref{subsec:delay_OA} for details. 
Our results here demonstrate that Eq.~\eqref{eq:dynamics_z_identical} 
obtained from the microscopic Kuramoto model with three-body interactions, Eq.~\eqref{eq:dynamics_second_identical},
is consistent with 
the macroscopic equation obtained by directly applying the OA ansatz to the
time-delayed Kuramoto model up to the second order.

\section{Concluding remarks}

We have demonstrated that effective higher-order interactions emerge from delayed pairwise interactions
in Kuramoto-type oscillator systems, and compared their behaviors both numerically and analytically.
Our results confirmed that, for moderately small $\epsilon$ and $\tau$, the delayed Kuramoto model and the 
three-body Kuramoto model exhibit 
qualitatively consistent synchronization transitions.
In particular, both models exhibited bistability between coherence and incoherence.
Our results show that time-delayed interactions can be interpreted in terms of higher-order interactions, while also suggesting the possibility of interpreting higher-order interactions in terms of time delay.

Generally, if we perform higher-order expansions to approximate a time-delay system, the number of terms required depends on the specific dynamical features and the desired level of accuracy. In this study, our primary focus was bistability; therefore, truncating the expansion at three-body couplings was qualitatively sufficient. However, if the target time-delayed system exhibits more complex dynamics, higher-order terms beyond three-body interactions may also be needed. Development of a systematic criterion for determining the required truncation order is an important open problem.
 
While we focused only on the globally coupled Kuramoto models using the OA ansatz for the analysis in this study, the effective higher-order interpretation is adaptable to other types of coupled oscillator networks, such as regular lattices or random networks,
facilitating detailed analysis and control design for a wide range of network systems involving time delays.

\section*{ACKNOWLEDGMENTS}
We thank C. Bick, K. Yawata and I. Le\'on for useful comments.
We also thank the anonymous reviewer for pointing out that the stability results can also be obtained by directly applying the OA ansatz to the original time-delayed Kuramoto model (Appendix \ref{subsec:delay_OA}).
N.F. acknowledges JST SPRING, grant JPMJSP2106 and JPMJSP2180. N.F. and H.N acknowledge JSPS KAKENHI 25H01468, 25K03081, 22H00516 and 22K11919. K.T acknowledges JSPS KAKENHI 24K20863.  R.M. acknowledges JSPS KAKENHI 24KF0211. 

\begin{widetext}

\appendix
\section{Derivation of the higher-order interactions from time-delayed pairwise coupling}
\label{subsec:derivation_higher-order}
We show the details of the calculation to derive the higher-order Kuramoto model, Eq.~(\ref{eq:dynamics_second}), from the time-delayed Kuramoto model, Eq.~(\ref{eq:Kuramoto_time_delay}), under the assumption of small $\epsilon$. 
Recall from Eq.~(\ref{eq:taylor_expansionchanged}) that 
{\small
\begin{align}\label{eq:thetaequationintegralappendix}
     \dot{\theta}_j(t)
    = \omega_j+\frac{\epsilon}{N}\sum_{\substack{k=1\\k\neq j}}^{N}
    \sin\left(\theta_k(t-\tau) -\theta_j(t)\right) = \omega_j+\frac{\epsilon}{N}\sum_{\substack{k=1\\k\neq j}}^{N}
    \sin\left(\theta_k(t) - \int_0^{\tau} \dot \theta_k(t-s)ds -\theta_j(t)\right)\, .
\end{align}
}
It follows in particular that 
\begin{align}\label{eq:thetadottmins}
\dot \theta_k(t-s) & = \omega_k + \frac{\epsilon}{N} \sum_{\substack{l=1\\l\neq k}}^N\sin\left(\theta_l(t-s) - \int_0^{\tau} \dot \theta_l(t-s-\sigma)d\sigma -\theta_k(t-s)\right)   \notag \\
& =  \omega_k + \frac{\epsilon}{N} \sum_{\substack{l=1\\l\neq k}}^N\sin\left(\theta_l(t) -\omega_l s - \omega_l \tau -\theta_k(t)+\omega_k s + \mathcal{O}(\epsilon) \right) \notag \\
& =  \omega_k + \frac{\epsilon}{N} \sum_{\substack{l=1\\l\neq k}}^N\sin\left(\theta_l(t) -\theta_k(t)  - \omega_l \tau +(\omega_k-\omega_l ) s  \right) + \mathcal{O}(\epsilon^2) \, , \notag \\
\end{align}
where the second equality holds because $\theta_l(t-s)=\theta_l(t)-\omega_ls + \mathcal{O}(\epsilon)$, $\dot \theta_l = \omega_l + \mathcal{O}(\epsilon)$, and $\theta_k(t-s)=\theta_k(t)-\omega_ks + \mathcal{O}(\epsilon)$, and the third equality follows from Taylor expanding the sinus. Substituting Eq.~(\ref{eq:thetadottmins}) into Eq.~(\ref{eq:thetaequationintegralappendix}), we find that
{\small
\begin{align}    \label{eq:thetadotwithintegral}
    \dot{\theta}_j(t)=&\ \omega_j+\frac{\epsilon}{N}\sum_{\substack{k=1\\k\neq j}}^{N}\sin\left[\theta_k(t)-\theta_j(t)
    -\omega_k\tau-\frac{\epsilon}{N}\sum_{\substack{l=1\\l\neq k}}^{N} \int_0^{\tau}\sin\left\{\theta_l(t) -\theta_k(t)-\omega_l\tau +(\omega_k-\omega_l)s\right\}  ds     \right] + \mathcal{O}(\epsilon^3) \notag \\
     =&\ \omega_j  +\frac{\epsilon}{N}\sum_{\substack{k=1\\k\neq j}}^{N}  \sin\left( \theta_k(t)-\theta_j(t) - \omega_k\tau \right) \notag \\  &  -\frac{\epsilon^2}{N^2}\sum_{\substack{k=1\\k\neq j}}^{N}\sum_{\substack{l=1\\l\neq k}}^{N} \cos\left( \theta_k(t)-\theta_j(t) - \omega_k\tau \right)\int_0^{\tau}\!\!\sin\left\{\theta_l(t) -\theta_k(t)-\omega_l\tau +(\omega_k-\omega_l)s\right\}  ds + \mathcal{O}(\epsilon^3)\, . 
\end{align}
}
Now it remains to note that the integrals in this expression evaluate to
$$\int_0^{\tau}\!\!\sin\left\{\theta_l(t) -\theta_k(t)-\omega_l\tau +(\omega_k-\omega_l)s\right\}  ds = \left\{ \begin{array}{ll} \frac{\cos(\theta_l(t)-\theta_k(t)-\omega_l\tau)-\cos(\theta_l(t)-\theta_k(t)+(\omega_k-2\omega_l)\tau)}{\omega_k-\omega_l}  & , \mbox{when} \ \omega_k\neq \omega_l \\ \tau\sin\left( \theta_l(t)-\theta_k(t) -\omega_l\tau \right)  & , \mbox{when} \ \omega_k= \omega_l  \end{array}\right. \, .$$
It is easy to verify that Eq.~(\ref{eq:thetadotwithintegral}) coincides with Eq.~(\ref{eq:dynamics_second}).

\section{Rewriting Eq.~(3) using the general complex order parameters}
\label{subsec:Eq8}
We derive Eq.~\eqref{eq:dynamics_second_order} as an approximate description 
of Eq.~\eqref{eq:dynamics_second} using the general complex order parameters under
the assumption of $\gamma=\mathcal{O}(\epsilon)$ and $N\gg 1$.
From Eq.~\eqref{eq:dynamics_second}, 
\begin{align}  
    \dot{\theta}_j(t)
    &=\omega_j+\frac{\epsilon}{N}\sum_{\substack{k=1\\k\neq j}}^{N}\frac{e^{i(\theta_k(t)-\theta_j(t)
    -\omega_k\tau)}-e^{-i(\theta_k(t)-\theta_j(t)
    -\omega_k\tau)}}{2i}\notag\\
    &\quad+\frac{\epsilon^2}{N^2}\sum_{\substack{k=1\\k\neq j}}^{N}\sum_{\substack{l=1\\l\neq k}}^{N}
    \frac{e^{i(\theta_k(t)-\theta_j(t)-\omega_k\tau)}+e^{-i(\theta_k(t)-\theta_j(t)-\omega_k\tau)}}{2}
    \left(\frac{e^{i(\theta_l(t)-\theta_k(t)-\omega_l\tau)}}{2}\frac{1-e^{-i(\omega_l-\omega_k)\tau}}{\omega_l-\omega_k}
    \right.\notag\\
    &\quad\left.
    +\frac{e^{-i(\theta_l(t)-\theta_k(t)-\omega_l\tau)}}{2}\frac{1-e^{i(\omega_l-\omega_k)\tau}}{\omega_l-\omega_k}\right).
\end{align}
Under the assumption that $\gamma=\mathcal{O}(\epsilon)$, we can suppose that $\omega_k-\omega_j=\mathcal{O}(\epsilon)$ holds for the majority of oscillator pairs $k$ and $j$. Consequently, $\frac{1- e^{\pm i(\omega_l-\omega_k)\tau}}{\omega_l-\omega_k}=\mp i\tau+\mathcal{O}(\epsilon)$. Hence, we obtain
{\small
\begin{align}      
    \dot{\theta}_j(t)&\simeq\omega_j+\frac{\epsilon}{N}\sum_{\substack{k=1\\k\neq j}}^{N}\frac{e^{i(\theta_k(t)-\theta_j(t)
    -\omega_k\tau)}-e^{-i(\theta_k(t)-\theta_j(t)
    -\omega_k\tau)}}{2i}\notag\\
    &\quad-\frac{\epsilon^2\tau}{N^2}\sum_{\substack{k=1\\k\neq j}}^{N}\sum_{\substack{l=1\\l\neq k}}^{N}
    \frac{e^{i(\theta_k(t)-\theta_j(t)-\omega_k\tau)}+e^{-i(\theta_k(t)-\theta_j(t)-\omega_k\tau)}}{2}
    \frac{e^{i(\theta_l(t)-\theta_k(t)-\omega_l\tau)}-e^{-i(\theta_l(t)-\theta_k(t)-\omega_l\tau)}}{2i}   \notag\\
    &=\omega_j+\frac{\epsilon}{N}\left\{\sum_{k=1}^{N}\frac{e^{i(\theta_k(t)-\theta_j(t)
    -\omega_k\tau)}-e^{-i(\theta_k(t)-\theta_j(t)
    -\omega_k\tau)}}{2i}-\frac{e^{-i\omega_j\tau}-e^{i\omega_j\tau}}{2i}\right\}\notag\\
    &\quad-\frac{\epsilon^2\tau}{N^2}\left[\sum_{k=1}^{N}
    \frac{e^{i(\theta_k(t)-\theta_j(t)-\omega_k\tau)}+e^{-i(\theta_k(t)-\theta_j(t)-\omega_k\tau)}}{2}\left\{\sum_{l=1}^{N}
    \frac{e^{i(\theta_l(t)-\theta_k(t)-\omega_l\tau)}-e^{-i(\theta_l(t)-\theta_k(t)-\omega_l\tau)}}{2i}-\frac{e^{-i\omega_k\tau}-e^{i\omega_k\tau}}{2i}\right\}\right.\notag\\
    &\quad-\left.
    \frac{e^{-i\omega_j\tau}+e^{i\omega_j\tau}}{2}\left\{\sum_{l=1}^{N}
    \frac{e^{i(\theta_l(t)-\theta_j(t)-\omega_l\tau)}-e^{-i(\theta_l(t)-\theta_j(t)-\omega_l\tau)}}{2i}-\frac{e^{-i\omega_j\tau}-e^{i\omega_j\tau}}{2i}\right\}\right].
 \end{align}
}%
\noindent Writing this in terms of the general complex order parameters defined in Eq.~(\ref{eq:order_parameter_with_lag}),
this becomes
\begin{align}       
    \dot{\theta}_j(t) 
    &\simeq\omega_j+\epsilon\left\{\frac{e^{-i\theta_j(t)}z_1(t)-e^{i\theta_j(t)}z_1^*(t)}{2i}+\mathcal{O}(N^{-1})\right\}\notag\\
    &\quad-\epsilon^2\tau\left[\sum_{k=1}^{N}
    \left\{\frac{e^{-i(\theta_j(t)+\omega_j\tau)}e^{i(\omega_j-\omega_k)\tau}z_1(t)-e^{-i(\theta_j(t)-\omega_j\tau)}e^{2i(\theta_k(t)-\omega_k\tau)}e^{-i(\omega_j-\omega_k)\tau}z_1^*(t)}{4iN}\right.\right.\notag\\
    &\quad\left.\left.+\frac{e^{i(\theta_j(t)-\omega_j\tau)}e^{-2i(\theta_k(t)-\omega_k\tau)}e^{i(\omega_j-\omega_k)\tau}z_1(t)-e^{i(\theta_j(t)+\omega_j\tau)}e^{-i(\omega_j-\omega_k)\tau}z_1^*(t)}{4iN})\right\}+\mathcal{O}(N^{-1})\right],
 \end{align}
where $e^{\pm i(\omega_j-\omega_k)\tau}=1+\mathcal{O}(\epsilon)$. Then,
{\small
\begin{align}   
    \dot{\theta}_j(t)&\simeq
    \omega_j+\epsilon\left\{\frac{e^{-i\theta_j(t)}z_1(t)-e^{i\theta_j(t)}z_1^*(t)}{2i}+\mathcal{O}(N^{-1})\right\}\notag\\
    &\quad-\epsilon^2\tau\left\{
    \frac{e^{-i(\theta_j(t)+\omega_j\tau)}z_1(t)-e^{-i(\theta_j(t)-\omega_j\tau)}z_1^*(t)z_2(t)+e^{i(\theta_j(t)-\omega_j\tau)}z_1(t)z_2^*(t)-e^{i(\theta_j(t)+\omega_j\tau)}z_1^*(t)}{4i}+\mathcal{O}(N^{-1})\right\}.
 \end{align}
}%
\noindent Under the assumption that $N\gg 1$, we therefore find
\begin{align}
    \dot{\theta}_j(t)
    \simeq
    \omega_j+\frac{1}{2i}\left\{e^{-i\theta_j(t)}H_j(t)-e^{i\theta_j(t)}H_j^*(t)\right\},
\end{align}
where $H_j=\epsilon z_1-\frac{\epsilon^2\tau}{2}\left(e^{-i\omega_j\tau}z_1-e^{i\omega_j\tau}z^*_{1}z_2\right)$.

\section{Ott-Antonsen ansatz}

\label{subsec:OA}
We derive the macroscopic description of Eq.~\eqref{eq:dynamics_second_order}
via the Ott-Antonsen(OA) ansatz.
In the $N\rightarrow \infty$ limit, the general complex order parameter in Eq.~\eqref{eq:order_parameter_with_lag} is redefined as
\begin{align}
    z_m(t)=\int_{0}^{2\pi}\int_{-\infty}^{\infty}e^{i m (\theta'-\omega'\tau)}
    P(\theta',~\omega',~t)g(\omega')d\omega'd\theta',
    \label{eq:order_parameter_with_P}
\end{align}
where $P(\theta,~\omega,~t)$ is the probability density function of the oscillator phase $\theta$ at time $t$ and the frequency $\omega$,
and the time evolution of $P(\theta,\omega, t)$ is given from Eq.~(\ref{eq:dynamics_second_order}) as
\begin{align}
    \frac{\partial P}{\partial t}&=-\frac{\partial}{\partial \theta} \left( \dot{\theta}P \right)\notag\\
    &=-\frac{\partial}{\partial \theta} \left[ \left\{
    \omega+\frac{1}{2i}\left(e^{-i\theta}H-e^{i\theta}
    H^*\right) \right\} P \right],
    \label{eq:dynamics_of_P}
\end{align}
where $H = \epsilon z_1-\frac{\epsilon^2\tau}{2}\left(e^{-i\omega \tau}z_1-
e^{i\omega \tau}z^*_{1}z_2\right)$.

Now, we expand $P(\theta,\omega, t)$ in Fourier series as
\begin{align}
P(\theta,\omega, t)=\frac{1}{2 \pi} \sum_{m=-\infty}^{\infty} P_m(\omega, t) e^{im\theta},
\label{eq:PDF_fourier}
\end{align}
where $P_m$ is the $m$-th Fourier coefficient.
Note that $P_0(\omega, t)=1$ by the normalization condition for $P(\theta,\omega, t)$, i.e., 
$\int_0^{2\pi} P(\theta, \omega, t) d\theta = 1$, and $P_m(\omega, t)={P}^*_{-m}(\omega, t)$ since $P(\theta,\omega, t)$ is real.
Substituting Eq.~(\ref{eq:PDF_fourier}) into Eq.~(\ref{eq:dynamics_of_P}), we obtain the equation for $m=1$ as
\begin{align}
    \frac{\partial P_1(\omega, t)}{\partial t} 
    &=-im\omega P_1(\omega, t) -\frac{1}{2}\left(
    P_2(\omega, t)G-
     G^*\right),
\end{align}
where
{\small
\begin{align}
G&=\left(\epsilon-\frac{\epsilon^2\tau}{2}e^{-i\omega\tau}\right)\int_{-\infty}^{\infty}
    e^{-i\omega'\tau}P_{-1}(\omega', t) g(\omega')d\omega'+\frac{\epsilon^2\tau}{2}
    e^{i\omega\tau}\int_{-\infty}^{\infty}e^{i\omega'\tau}P_1(\omega', t) 
    g(\omega')d\omega'\int_{-\infty}^{\infty}e^{-2i \omega'\tau}P_{-2}(\omega', t) 
    g(\omega')d\omega'.
\end{align}
}%
\noindent We now introduce the OA ansatz and assume that the Fourier coefficient $P_m$ can be represented as
\begin{align}
        P_m(\omega,~t)=\alpha(\omega,~t)^m,~~(m=1,~2,~\cdots).
        \label{eq:Ott-Antonsen_ansatz}
\end{align}
The coefficient $\alpha(\omega, t)$ then obeys
\begin{align}
    \frac{\partial \alpha(\omega,~t)}{\partial t}&=-i\omega \alpha(\omega,~t)-\frac{1}{2}\left[\alpha(\omega,~t)^2\left\{\left(\epsilon-
    \frac{\epsilon^2\tau}{2}e^{-i\omega\tau}\right)z_1(t)+\frac{\epsilon^2\tau}{2}
    e^{i\omega\tau}{z}^*_1(t)z_2(t)\right\}\right.\notag\\
    &\quad\left.- \left\{\left(\epsilon-\frac{\epsilon^2\tau}{2}e^{i\omega\tau}\right){z}^*_1(t)+\frac{\epsilon^2\tau}{2}e^{-i\omega\tau}z_1(t){z}^*_2(t)\right\}\right],
    \label{eq:dynamics_alpha}
\end{align}
where
\begin{align}
        z_1(t)&=\int_{-\infty}^{\infty}\alpha^*(\omega',~t)
        e^{-i\omega'\tau}g(\omega')d\omega',\\
        z_2(t)&=\int_{-\infty}^{\infty}\left\{\alpha^*(\omega',~t)\right\}^2
        e^{-2i\omega'\tau}g(\omega')d\omega',
\end{align}
from Eq.~(\ref{eq:order_parameter_with_P}) and (\ref{eq:Ott-Antonsen_ansatz}).
Under the assumption that $g(\omega)$  follows the Lorentzian distribution, we obtain
\begin{align}
    z_1(t)
    &={\alpha^*(\omega_0-i\gamma,~t)}e^{-i(\omega_0-i\gamma)\tau},\\
    z_2(t)
    &=\left\{\alpha^*(\omega_0-i\gamma,~t)\right\}^2e^{-2i(\omega_0-i\gamma)\tau}=\left\{z_1(t)\right\}^2,
\end{align}
by applying the residue theorem, where
$\alpha(\omega,~t)$ is analytically continued to the complex plane.
Then, Eq.~\eqref{eq:dynamics_alpha} is rewritten as
{\small
\begin{align}
    \dot{z_1}(t)
    &=\left\{i\omega_0-\gamma+ \frac{\epsilon}{2}e^{-i(\omega_0+i\gamma)\tau}-
    \frac{\epsilon^2\tau}{4}
    e^{-2i(\omega_0+i\gamma)\tau}-\left(\frac{\epsilon}{2}e^{i(\omega_0+i\gamma)\tau}-\frac{\epsilon^2\tau}{4}-\frac{\epsilon^2\tau}{4}e^{2i(\omega_0+i\gamma)\tau}
    \right)|{z}_1(t)|^2-\frac{\epsilon^2\tau}{4}
    |{z}_1(t)|^4\right\}{z}_1(t).
    \label{eq:dynamics_z}
\end{align}
}%
\noindent For the case of identical oscillators, we consider the limit with $\gamma\rightarrow +0$ and derive
\begin{align}
    \dot{z_1}(t)
    &=\left\{i\omega_0+ \frac{\epsilon}{2}e^{-i\omega_0\tau}-
    \frac{\epsilon^2\tau}{4}
    e^{-2i\omega_0\tau}
    -\left(\frac{\epsilon}{2}e^{i\omega_0\tau}-\frac{\epsilon^2\tau}{4}-\frac{\epsilon^2\tau}{4}e^{2i\omega_0\tau}
    \right)|{z}_1(t)|^2-\frac{\epsilon^2\tau}{4}
    |{z}_1(t)|^4\right\}{z}_1(t),
    \label{eq:dynamics_z1_identical}
\end{align}
where $z_1$ is redefined in the $N\rightarrow\infty$ limit and $\gamma\rightarrow +0$ limit as:
\begin{equation}
    z_1(t)=\int_{0}^{2\pi}\int_{-\infty}^{\infty}e^{i(\theta'-\omega'\tau)}
    P(\theta',~\omega',~t)\delta(\omega'-\omega_0)d\omega'd\theta'=e^{-i\omega_0\tau}\int_{0}^{2\pi}e^{i\theta'}
    P(\theta',~\omega_0,~t)d\theta'=e^{-i\omega_0\tau}z(t),
    \label{eq:z1_identical_continuum_limit}
\end{equation}
where the conventional complex order parameter is also  redefined in the $N\rightarrow\infty$ limit as $z(t)=\int_{0}^{2\pi}e^{i\theta'}
    P(\theta',~\omega_0,~t)d\theta'$.
It is easy to derive Eq.~\eqref{eq:dynamics_z_identical} from 
Eq.~\eqref{eq:dynamics_z1_identical} and \eqref{eq:z1_identical_continuum_limit}.
Note that within the OA ansatz, not all solutions are attractive for identical oscillators~\cite{pikovsky2011wsansatz}; however, this point is beyond the scope of the present work and is therefore not addressed here.

\section{Linear stability analysis of the complex order parameter dynamics}
\label{subsec:linear_stability_OA}
We implement a linear stability analysis for 
the complex order parameter dynamics derived by the OA ansatz in Eq.~\eqref{eq:dynamics_z_identical}.
Eq.~\eqref{eq:dynamics_z_identical} can be rewritten as
\begin{align}
    \dot{z}(t)
    &=\left\{i\omega_0+ \frac{\epsilon}{2}e^{-i\omega_0\tau}-
    \frac{\epsilon^2\tau}{4}
    e^{-2i\omega_0\tau}
    -\left(\frac{\epsilon}{2}e^{i\omega_0\tau}-\frac{\epsilon^2\tau}{4}-\frac{\epsilon^2\tau}{4}e^{2i\omega_0\tau}
    \right)z(t){z}^*(t)-\frac{\epsilon^2\tau}{4}
    \left\{z(t)\right\}^2\left\{z^*(t)\right\}^2\right\}{z}(t)=F\left(z,{z}^*\right).
\end{align}
This system possesses a trivial fixed point at $z=0$ and a limit cycle 
$z(t)=e^{i\Omega_s t}$ with $\Omega_s=\omega_0-\epsilon\sin\left(\omega_0\tau\right)+
\frac{\epsilon^2\tau}{2}\sin\left(2\omega_0\tau\right)$ independently of the parameter values.
Another intermediate limit cycle 
$z(t)=\tilde{R}e^{i\Omega_i t}$ with
$\tilde{R}=\sqrt{-\frac{2}{\epsilon\tau}\cos(\omega_0 \tau)+\cos(2\omega_0 \tau)}$ and
$\Omega_i=\omega_0 -\left(\frac{\epsilon}{2}\sin(\omega_0 \tau)-\frac{\epsilon^2\tau}{4}\sin(2\omega_0\tau)\right)\left(1+{\tilde{R}^2}\right)$,
also exists when $0\leq -\frac{2}{\epsilon\tau}\cos(\omega_0 \tau)+\cos(2\omega_0 \tau)\leq 1$.

We analyze the linearized equation described as
\begin{align}
    \begin{pmatrix}
    \dot{y}(t) \\
    \dot{y}^*(t)
    \end{pmatrix}=
    \left.
    \begin{pmatrix}
    \frac{\partial F\left(z,z^*\right)}{\partial z} & \frac{\partial F\left(z,z^*\right)}{\partial z^*} \\
    \frac{\partial {F^*\left(z,z^*\right)}}{\partial z} & \frac{\partial {F^*\left(z,z^*\right)}}{\partial z^*}
    \end{pmatrix}
    \right|_{\substack{z=z(t)\\z^*=z^*(t)}}
    \begin{pmatrix}
    {y}(t) \\
    {y^*}(t)
    \end{pmatrix},
    \label{eq:linearized_z}
\end{align}
where
\begin{align}
    \frac{\partial F\left(z,{z}^*\right)}{\partial z}
    &=i\omega_0+ \frac{\epsilon}{2}e^{-i\omega_0\tau}-
    \frac{\epsilon^2\tau}{4}
    e^{-2i\omega_0\tau}
    -\left(\epsilon e^{i\omega_0\tau}-\frac{\epsilon^2\tau}{2}-\frac{\epsilon^2\tau}{2}e^{2i\omega_0\tau}
    \right)z{z}^*-\frac{\epsilon^2\tau}{4}
    3z^2(z^*)^2,\\
    \frac{\partial F\left(z,{z}^*\right)}{\partial {z}^*}
    &=
    -\left(\frac{\epsilon}{2}e^{i\omega_0\tau}-\frac{\epsilon^2\tau}{4}-\frac{\epsilon^2\tau}{4}e^{2i\omega_0\tau}
    \right)z^2-\frac{\epsilon^2\tau}{2}
    z^3{z}^*,\\
    \frac{\partial {F^*\left(z,{z}^*\right)}}{\partial z}
    &=
    -\left(\frac{\epsilon}{2}e^{-i\omega_0\tau}-\frac{\epsilon^2\tau}{4}-\frac{\epsilon^2\tau}{4}e^{-2i\omega_0\tau}
    \right)(z^*)^2-\frac{\epsilon^2\tau}{2}
    z(z^*)^3,\\
    \frac{\partial {F^*\left(z,{z}^*\right)}}{\partial {z}^*}
    &=-i\omega_0+ \frac{\epsilon}{2}e^{i\omega_0\tau}-
    \frac{\epsilon^2\tau}{4}
    e^{2i\omega_0\tau}
    -\left(\epsilon e^{-i\omega_0\tau}-\frac{\epsilon^2\tau}{2}-\frac{\epsilon^2\tau}{2}e^{-2i\omega_0\tau}
    \right)z{z}^*-\frac{\epsilon^2\tau}{4}
    3z^2(z^*)^2.
\end{align}
For convenience, we introduce a rotating coordinate as $u(t)=y(t)e^{-i\Omega t}$. 
Then, Eq.~\eqref{eq:linearized_z} can be rewritten as
\begin{align}
    \begin{pmatrix}
    \dot{u}(t) \\
    \dot{{u}^*(t)}
    \end{pmatrix}=
    \left.
    \begin{pmatrix}
    \frac{\partial F\left(z,{z}^*\right)}{\partial z}-i\Omega & \frac{\partial F\left(z,{z}^*\right)}{\partial {z}^*}e^{-2i\Omega\tau} \\
    \frac{\partial {F^*\left(z,{z}^*\right)}}{\partial z}e^{2i\Omega\tau} & \frac{\partial {F^*\left(z,{z}^*\right)}}{\partial {z}^*}+i\Omega
    \end{pmatrix}
    \right|_{\substack{z=z(t)\\z^*=z^*(t)}}
    \begin{pmatrix}
    {u}(t) \\
    {{u}^*(t)}
    \end{pmatrix}.
\end{align}

For the fixed point at $z=0$, we obtain
\begin{align}
    \left.
    \begin{pmatrix}
    \frac{\partial F\left(z,{z}^*\right)}{\partial z}-i\Omega & \frac{\partial F\left(z,{z}^*\right)}{\partial {z}^*}e^{-2i\Omega\tau} \\
    \frac{\partial {F^*\left(z,{z}^*\right)}}{\partial z}e^{2i\Omega\tau} & \frac{\partial {F^*\left(z,{z}^*\right)}}{\partial {z}^*}+i\Omega
    \end{pmatrix}
    \right|_{\substack{z=0\\z^*=0}}
    =
    \begin{pmatrix}
     \frac{\epsilon}{2}e^{-i\omega_0\tau}-
        \frac{\epsilon^2\tau}{4}e^{-2i\omega_0\tau}&0\\
    0& \frac{\epsilon}{2}e^{i\omega_0\tau}-
        \frac{\epsilon^2\tau}{4}e^{2i\omega_0\tau}
    \end{pmatrix},
\end{align}
where we set $\Omega=\omega_0$. The corresponding eigenvalues are
\begin{align}
    \lambda=\frac{\epsilon}{2}e^{\mp i\omega_0\tau}-
    \frac{\epsilon^2\tau}{4}e^{\mp 2i\omega_0\tau}.
    \label{eq:eigen_R0}
\end{align}
The fixed point at $z=0$ is linearly stable under the condition $\frac{\epsilon\tau}{2}\cos(2\omega_0 \tau)>\cos(\omega_0 \tau)$.

For the limit cycle $z(t)=e^{i\Omega_s t}$, we find
\begin{align}
    &\left.
    \begin{pmatrix}
    \frac{\partial F\left(z,{z}^*\right)}{\partial z}-i\Omega & \frac{\partial F\left(z,{z}^*\right)}{\partial {z}^*}e^{-2i\Omega\tau} \\
    \frac{\partial {F^*\left(z,{z}^*\right)}}{\partial z}e^{2i\Omega\tau} & \frac{\partial {F^*\left(z,{z}^*\right)}}{\partial {z}^*}+i\Omega
    \end{pmatrix}
    \right|_{\substack{z=e^{i\Omega_s t}\\z^*=e^{-i\Omega_s t}}}
    =-
    \begin{pmatrix}
    \frac{\epsilon}{2} e^{i\omega_0\tau}+\frac{\epsilon^2\tau}{4}-\frac{\epsilon^2\tau}{4}e^{2i\omega_0\tau}
    &\frac{\epsilon}{2}e^{i\omega_0\tau}+\frac{\epsilon^2\tau}{4}-\frac{\epsilon^2\tau}{4}e^{2i\omega_0\tau}
    \\
    \frac{\epsilon}{2}e^{-i\omega_0\tau}+\frac{\epsilon^2\tau}{4}-\frac{\epsilon^2\tau}{4}e^{-2i\omega_0\tau}
    &\frac{\epsilon}{2} e^{-i\omega_0\tau}+\frac{\epsilon^2\tau}{4}-\frac{\epsilon^2\tau}{4}e^{-2i\omega_0\tau}
    \end{pmatrix},
\end{align}
where we set $\Omega=\Omega_s$
and the eigenvalues 
\begin{align}
    \lambda=0,~-\epsilon\cos\left(\omega_0\tau\right)+\frac{\epsilon^2\tau}{2}\left(\cos\left(2\omega_0\tau\right)-1\right).
    \label{eq:eigen_R1}
\end{align}
The limit cycle $z=e^{i\Omega_s t}$ is asymptotically stable under the condition $\frac{\epsilon\tau}{2}\left(\cos(2\omega_0 \tau)-1\right)<\cos(\omega_0 \tau)$.

For the intermediate limit cycle  $z(t)=\tilde{R}e^{i \Omega_i t}$,
we obtain
\begin{align}
    &\left.
    \begin{pmatrix}
    \frac{\partial F\left(z,{z}^*\right)}{\partial z}-i\Omega & \frac{\partial F\left(z,{z}^*\right)}{\partial {z}^*}e^{-2i\Omega\tau} \\
    \frac{\partial {F^*\left(z,{z}^*\right)}}{\partial z}e^{2i\Omega\tau} & \frac{\partial {F^*\left(z,{z}^*\right)}}{\partial {z}^*}+i\Omega
    \end{pmatrix}
    \right|_{\substack{z=\tilde{R}e^{i\Omega_i t}\\z^*=\tilde{R}e^{-i\Omega_i t}}}\notag\\
    &~~~~~~~~~~~~~~~~~~~=-\tilde{R}^2
    \begin{pmatrix}
        \frac{\epsilon}{2} e^{i\omega_0\tau}-\frac{\epsilon^2\tau}{4}-\frac{\epsilon^2\tau}{4}e^{2i\omega_0\tau}+\frac{\epsilon^2\tau}{2}
        \tilde{R}^2&\frac{\epsilon}{2}e^{i\omega_0\tau}-\frac{\epsilon^2\tau}{4}-\frac{\epsilon^2\tau}{4}e^{2i\omega_0\tau}
        +\frac{\epsilon^2\tau}{2}\tilde{R}^2
        \\\frac{\epsilon}{2}e^{-i\omega_0\tau}-\frac{\epsilon^2\tau}{4}-\frac{\epsilon^2\tau}{4}e^{-2i\omega_0\tau}
        +\frac{\epsilon^2\tau}{2}\tilde{R}^2
        &\frac{\epsilon}{2} e^{-i\omega_0\tau}-\frac{\epsilon^2\tau}{4}-\frac{\epsilon^2\tau}{4}e^{-2i\omega_0\tau}
        +\frac{\epsilon^2\tau}{2}\tilde{R}^2
    \end{pmatrix},
\end{align}
where we set $\Omega=\Omega_i$ and the eigenvalues 
\begin{align}
    \lambda=0,~\frac{\epsilon^2\tau}{2}\tilde{R}^2\left(1-\tilde{R}^2\right).
\end{align}
It is clear that $z(t)=\tilde{R}e^{i\tilde\Omega_i t}$ cannot be an asymptotically stable limit cycle.

\section{Derivation of Eq.~(10) from the delay-differential equation 
of the complex order parameter}
\label{subsec:delay_OA}

We show that the equation for the complex order parameter, Eq.~\eqref{eq:dynamics_z_identical}, which is derived by applying 
the OA ansatz to the three-body Kuramoto model, Eq.~\eqref{eq:dynamics_second_identical}, 
and the stability of the incoherent state and the fully synchronized state 
can also be derived from the delay-differential equation 
of the complex order parameter
derived by directly applying the OA ansatz to the time-delayed pairwise Kuramoto model, Eq.~\eqref{eq:Kuramoto_time_delay}.
As shown in \cite{Ott_Antonsen_2008, Ott_Antonsen_2009}, 
the delay-differential equation for the complex order parameter of 
the time-delayed Kuramoto model is
\begin{align}
    \dot{z}(t)=i\omega_0z(t)+\frac{\epsilon}{2}\left\{ 
    z(t-\tau)- z^*(t-\tau)\left(z(t)\right)^2\right\}.
     \label{eq:delay_OA}
\end{align}
For convenience, we introduce a rotating coordinate such that $z = \hat{z}e^{i\omega_0t}$. Then, we obtain
\begin{align}
    &\dot{\hat{z}}(t)
    =\frac{\epsilon}{2}\left\{e^{-i\omega_0\tau}
    \hat{z}(t-\tau)- e^{i\omega_0\tau}\hat{z}^*(t-\tau)
    \left(\hat{z}(t)\right)^2\right\}.
    \label{eq:delay_OA_rotating}
\end{align}
For the delay term, we expand as
\begin{align}
    \hat{z}(t-\tau) &= \hat{z}(t) -\tau\dot{\hat{z}}(t) +\cdots,\notag\\
    &= \hat{z}(t) -\frac{\epsilon\tau}{2}\left\{e^{-i\omega_0\tau}
    \hat{z}(t-\tau)- e^{i\omega_0\tau}\hat{z}^*(t-\tau)
    \left(\hat{z}(t)\right)^2\right\} + \mathcal{O}(\epsilon^2),\notag\\
    &= \left[1 -\frac{\epsilon\tau}{2}\left\{e^{-i\omega_0\tau}
    - e^{i\omega_0\tau}|\hat{z}(t)|^2\right\}\right]\hat{z}(t) + \mathcal{O}(\epsilon^2).
\end{align}
Substituting into Eq.~\eqref{eq:delay_OA_rotating}, we obtain
\begin{align}
    &\dot{\hat{z}}(t)=\left\{
    \frac{\epsilon}{2}e^{-i\omega_0\tau} -
    \frac{\epsilon^2\tau}{4}e^{-2i\omega_0\tau}
    - \left( \frac{\epsilon}{2}e^{i\omega_0\tau}-\frac{\epsilon^2\tau}{4} -
    \frac{\epsilon^2\tau}{4}e^{2i\omega_0\tau}
    \right)|\hat{z}(t)|^2- \frac{\epsilon^2\tau}{4}|\hat{z}(t)|^4
    \right\}\hat{z}(t).
    \label{eq:delay_OA_rotating_2nd}
\end{align}
It is easy to verify that Eq.~\eqref{eq:delay_OA_rotating_2nd} for the original coordinate 
$z$ coincides with Eq.~\eqref{eq:dynamics_z_identical}.

The linear stability of the incoherent state $z=0$ and the fully synchronized state $z=e^{i\Omega t}$ for 
Eq.~\eqref{eq:dynamics_z_identical} can also be derived from Eq.~\eqref{eq:delay_OA} by approximation.
First, the stability of $z=0$ for Eq.~\eqref{eq:delay_OA_rotating} is derived as follows.
The linearized delay-differential equation of Eq.~\eqref{eq:delay_OA_rotating} around
$\hat{z}=0$ is
\begin{align}
    \begin{pmatrix}
    \dot{u}(t) \\
    \dot{u^*}(t)
    \end{pmatrix}=
    \left.\frac{\epsilon}{2}
    \begin{pmatrix}
    e^{-i\omega_0 \tau} & 0\\
    0 & e^{i\omega_0 \tau}
    \end{pmatrix}
    \right|_{\substack{\hat{z}=0\\\hat{z}^*=0}}
    \begin{pmatrix}
    u(t-\tau) \\
    u^*(t-\tau)
    \end{pmatrix}.
    \label{eq:delay_OA_linear_incoherent}
\end{align}
We consider solutions of the form
\begin{align}
    u(t) = u_1 e^{\lambda t}, \quad
    u^*(t) = u_2 e^{\lambda t},
    \label{eq:pertubation}
\end{align}
where $\lambda \in \mathbb{C}$.
Substituting into Eq.~\eqref{eq:delay_OA_linear_incoherent}, we obtain a transcendental characteristic equation
\begin{align*}
\lambda=\frac{\epsilon}{2}e^{-(\lambda\pm i\omega_0)\tau}.
\end{align*}
Assuming $\lambda=\mathcal{O}(\epsilon)$, 
$\lambda\simeq\frac{\epsilon}{2}e^{\mp i\omega_0\tau}\left(1-\lambda\tau\right)
+\mathcal{O}(\epsilon^3)$.
Then, we obtain the approximation for $\lambda$ as
\begin{align}
    \lambda
    &\simeq\frac{\epsilon}{2}e^{\mp i\omega_0\tau}\left(1-\frac{\epsilon\tau}{2}e^{\mp i\omega_0\tau}\right).
\end{align}
These results are consistent with the stability of the incoherent state
for Eq.~\eqref{eq:dynamics_z_identical} in Eq.~\eqref{eq:eigen_R0}.

The stability of $z=e^{i\Omega t}$ for Eq.~\eqref{eq:delay_OA} derived as follows.
For convenience, we introduce another rotating coordinate such that $z = \bar{z}e^{i\Omega t}$,
here $\Omega = \omega_0 -\epsilon\sin\Omega\tau$, then we obtain
\begin{align}
    &\dot{\bar{z}}(t)
    =i\epsilon\sin(\Omega\tau)\bar{z}(t)+\frac{\epsilon}{2}\left\{
    \bar{z}(t-\tau)e^{-i\Omega \tau}-\bar{z}^*(t-\tau)
    e^{i\Omega \tau}\left(\bar{z}(t)\right)^2\right\}.
    \label{eq:delay_OA_rotating2}
\end{align}
The fully synchronized state is $\bar{z}=1$,
and the linearized delay-differential equation of 
Eq.~\eqref{eq:delay_OA_rotating2} around it is given as
\begin{align}
    \begin{pmatrix}
    \dot{u}(t) \\
    \dot{u^*}(t)
    \end{pmatrix}=
    \left.
    \begin{pmatrix}
    i\epsilon\sin(\Omega\tau)-\epsilon e^{i\Omega \tau} & 0\\
    0 & -i\epsilon\sin(\Omega\tau)-\epsilon e^{-i\Omega \tau}
    \end{pmatrix}
    \right|_{\substack{\bar{z}=1\\\bar{z}^*=1}}
    \begin{pmatrix}
    {u}(t) \\
    {u^*(t)}
    \end{pmatrix}
    + \left.
    \begin{pmatrix}
    \frac{\epsilon}{2}e^{-i\Omega \tau} & -\frac{\epsilon}{2}e^{i\Omega \tau}\\
    -\frac{\epsilon}{2}e^{-i\Omega \tau} & \frac{\epsilon}{2}e^{i\Omega \tau}
    \end{pmatrix}
    \right|_{\substack{\bar{z}=1\\\bar{z}^*=1}}
    \begin{pmatrix}
    {u}(t-\tau) \\
    {u^*(t-\tau)}
    \end{pmatrix}.
\end{align}
Substituting Eq.~\eqref{eq:pertubation} into the linearized equation 
yields
\begin{align}
    \lambda
    \begin{pmatrix}
    u_1 \\
    u_2
    \end{pmatrix}
    =
    M(\lambda)
    \begin{pmatrix}
    u_1 \\
    u_2
    \end{pmatrix},
\end{align}
where
\begin{align}
    M(\lambda)
    =
    \begin{pmatrix}
    i\epsilon\sin(\Omega\tau)-\epsilon e^{i\Omega \tau}
    +\frac{\epsilon}{2}e^{-i\Omega \tau}e^{-\lambda\tau}& -\frac{\epsilon}{2}e^{i\Omega \tau}
    e^{-\lambda\tau}\\
    -\frac{\epsilon}{2}e^{-i\Omega \tau}e^{-\lambda\tau} & -i\epsilon\sin(\Omega\tau)-\epsilon e^{-i\Omega \tau}+\frac{\epsilon}{2}e^{i\Omega \tau}
    e^{-\lambda\tau}
    \end{pmatrix}.
\end{align} 
The characteristic equation is therefore given by
$\det\left(M(\lambda) - \lambda I\right)=0$, which is transcendental.
Assuming $\lambda=\mathcal{O}(\epsilon)$, 
\begin{align}
M(\lambda)
=
-\frac{\epsilon}{2}
\begin{pmatrix}
e^{i\Omega\tau} &e^{i\Omega\tau}\\
e^{-i\Omega\tau}&e^{-i\Omega\tau} 
\end{pmatrix}
+ O(\epsilon^2).
\end{align}
Then, we obtain the following approximation for $\lambda$:
\begin{align}
    \lambda \simeq 0, \quad \lambda \simeq -\epsilon\cos(\Omega\tau).
\end{align}
Furthermore, substituting $\Omega\simeq\omega_0 -\epsilon
\sin(\omega_0\tau)-\frac{\epsilon^2\tau}{2}\sin(2\omega_0\tau)$ yields
\begin{align}
    \lambda \simeq 0, \quad \lambda \simeq -\epsilon
    \cos(\omega_0\tau)-\epsilon^2\tau\left(\sin(\omega_0\tau)\right)^2.
\end{align}
These are consistent with the stability of the fully synchronized state
for Eq.~\eqref{eq:dynamics_z_identical} in Eq.~\eqref{eq:eigen_R1}.

Note that while Eq.~\eqref{eq:delay_OA} 
has a multitude of synchronized states~\cite{choi2000delay, strogatz2003delay},
depending on the number of solutions of $\Omega = \omega_0 -\epsilon\sin\Omega\tau$, 
Eq.~\eqref{eq:dynamics_z_identical} has a unique fully synchronized state 
with $\Omega_s=\omega_0-\epsilon\sin\left(\omega_0\tau\right)+
\frac{\epsilon^2\tau}{2}\sin\left(2\omega_0\tau\right)$. 
Thus, the blue regions in Fig.~\ref{fig:stability_zoom}(a) indicate the existence of a unique stable fully synchronized state,
whereas the blue regions in Fig.~\ref{fig:stability_zoom}(b) indicate the existence of one or more stable fully synchronized states.

\end{widetext}


\begin{thebibliography}{64}%
\makeatletter
\providecommand \@ifxundefined [1]{%
 \@ifx{#1\undefined}
}%
\providecommand \@ifnum [1]{%
 \ifnum #1\expandafter \@firstoftwo
 \else \expandafter \@secondoftwo
 \fi
}%
\providecommand \@ifx [1]{%
 \ifx #1\expandafter \@firstoftwo
 \else \expandafter \@secondoftwo
 \fi
}%
\providecommand \natexlab [1]{#1}%
\providecommand \enquote  [1]{``#1''}%
\providecommand \bibnamefont  [1]{#1}%
\providecommand \bibfnamefont [1]{#1}%
\providecommand \citenamefont [1]{#1}%
\providecommand \href@noop [0]{\@secondoftwo}%
\providecommand \href [0]{\begingroup \@sanitize@url \@href}%
\providecommand \@href[1]{\@@startlink{#1}\@@href}%
\providecommand \@@href[1]{\endgroup#1\@@endlink}%
\providecommand \@sanitize@url [0]{\catcode `\\12\catcode `\$12\catcode `\&12\catcode `\#12\catcode `\^12\catcode `\_12\catcode `\%12\relax}%
\providecommand \@@startlink[1]{}%
\providecommand \@@endlink[0]{}%
\providecommand \url  [0]{\begingroup\@sanitize@url \@url }%
\providecommand \@url [1]{\endgroup\@href {#1}{\urlprefix }}%
\providecommand \urlprefix  [0]{URL }%
\providecommand \Eprint [0]{\href }%
\providecommand \doibase [0]{http://dx.doi.org/}%
\providecommand \selectlanguage [0]{\@gobble}%
\providecommand \bibinfo  [0]{\@secondoftwo}%
\providecommand \bibfield  [0]{\@secondoftwo}%
\providecommand \translation [1]{[#1]}%
\providecommand \BibitemOpen [0]{}%
\providecommand \bibitemStop [0]{}%
\providecommand \bibitemNoStop [0]{.\EOS\space}%
\providecommand \EOS [0]{\spacefactor3000\relax}%
\providecommand \BibitemShut  [1]{\csname bibitem#1\endcsname}%
\let\auto@bib@innerbib\@empty
\bibitem [{\citenamefont {Battiston}\ \emph {et~al.}(2020)\citenamefont {Battiston}, \citenamefont {Cencetti}, \citenamefont {Iacopini}, \citenamefont {Latora}, \citenamefont {Lucas}, \citenamefont {Patania}, \citenamefont {Young},\ and\ \citenamefont {Petri}}]{battiston2020networks}%
  \BibitemOpen
  \bibfield  {author} {\bibinfo {author} {\bibfnamefont {F.}~\bibnamefont {Battiston}}, \bibinfo {author} {\bibfnamefont {G.}~\bibnamefont {Cencetti}}, \bibinfo {author} {\bibfnamefont {I.}~\bibnamefont {Iacopini}}, \bibinfo {author} {\bibfnamefont {V.}~\bibnamefont {Latora}}, \bibinfo {author} {\bibfnamefont {M.}~\bibnamefont {Lucas}}, \bibinfo {author} {\bibfnamefont {A.}~\bibnamefont {Patania}}, \bibinfo {author} {\bibfnamefont {J.-G.}\ \bibnamefont {Young}}, \ and\ \bibinfo {author} {\bibfnamefont {G.}~\bibnamefont {Petri}},\ }\href@noop {} {\bibfield  {journal} {\bibinfo  {journal} {Phys. Rep.}\ }\textbf {\bibinfo {volume} {874}},\ \bibinfo {pages} {1} (\bibinfo {year} {2020})}\BibitemShut {NoStop}%
\bibitem [{\citenamefont {Bianconi}(2021)}]{bianconi2021higher}%
  \BibitemOpen
  \bibfield  {author} {\bibinfo {author} {\bibfnamefont {G.}~\bibnamefont {Bianconi}},\ }\href@noop {} {\emph {\bibinfo {title} {Higher-{O}rder {N}etworks: {A}n introduction to simplicial complexes}}}\ (\bibinfo  {publisher} {Cambridge {U}niversity {P}ress},\ \bibinfo {year} {2021})\BibitemShut {NoStop}%
\bibitem [{\citenamefont {Battiston}\ \emph {et~al.}(2021)\citenamefont {Battiston}, \citenamefont {Amico}, \citenamefont {Barrat}, \citenamefont {Bianconi}, \citenamefont {de~Arruda}, \citenamefont {Franceschiello}, \citenamefont {Iacopini}, \citenamefont {Kéfi}, \citenamefont {Latora}, \citenamefont {Moreno}, \citenamefont {Murray}, \citenamefont {Peixoto}, \citenamefont {Vaccarino},\ and\ \citenamefont {Petri}}]{natphys}%
  \BibitemOpen
  \bibfield  {author} {\bibinfo {author} {\bibfnamefont {F.}~\bibnamefont {Battiston}}, \bibinfo {author} {\bibfnamefont {E.}~\bibnamefont {Amico}}, \bibinfo {author} {\bibfnamefont {A.}~\bibnamefont {Barrat}}, \bibinfo {author} {\bibfnamefont {G.}~\bibnamefont {Bianconi}}, \bibinfo {author} {\bibfnamefont {G.}~\bibnamefont {de~Arruda}}, \bibinfo {author} {\bibfnamefont {B.}~\bibnamefont {Franceschiello}}, \bibinfo {author} {\bibfnamefont {I.}~\bibnamefont {Iacopini}}, \bibinfo {author} {\bibfnamefont {S.}~\bibnamefont {Kéfi}}, \bibinfo {author} {\bibfnamefont {V.}~\bibnamefont {Latora}}, \bibinfo {author} {\bibfnamefont {Y.}~\bibnamefont {Moreno}}, \bibinfo {author} {\bibfnamefont {M.}~\bibnamefont {Murray}}, \bibinfo {author} {\bibfnamefont {T.}~\bibnamefont {Peixoto}}, \bibinfo {author} {\bibfnamefont {F.}~\bibnamefont {Vaccarino}}, \ and\ \bibinfo {author} {\bibfnamefont {G.}~\bibnamefont {Petri}},\ }\href@noop {} {\bibfield  {journal} {\bibinfo  {journal} {Nat. Phys.}\ }\textbf {\bibinfo {volume}
  {17}},\ \bibinfo {pages} {1093} (\bibinfo {year} {2021})}\BibitemShut {NoStop}%
\bibitem [{\citenamefont {Majhi}\ \emph {et~al.}(2022)\citenamefont {Majhi}, \citenamefont {Perc},\ and\ \citenamefont {Ghosh}}]{Dibakar_2022}%
  \BibitemOpen
  \bibfield  {author} {\bibinfo {author} {\bibfnamefont {S.}~\bibnamefont {Majhi}}, \bibinfo {author} {\bibfnamefont {M.}~\bibnamefont {Perc}}, \ and\ \bibinfo {author} {\bibfnamefont {D.}~\bibnamefont {Ghosh}},\ }\href {\doibase 10.1098/rsif.2022.0043} {\bibfield  {journal} {\bibinfo  {journal} {Journal of The Royal Society Interface}\ }\textbf {\bibinfo {volume} {19}},\ \bibinfo {pages} {20220043} (\bibinfo {year} {2022})}\BibitemShut {NoStop}%
\bibitem [{\citenamefont {Bick}\ \emph {et~al.}(2023)\citenamefont {Bick}, \citenamefont {Gross}, \citenamefont {Harrington},\ and\ \citenamefont {Schaub}}]{bick2023higher}%
  \BibitemOpen
  \bibfield  {author} {\bibinfo {author} {\bibfnamefont {C.}~\bibnamefont {Bick}}, \bibinfo {author} {\bibfnamefont {E.}~\bibnamefont {Gross}}, \bibinfo {author} {\bibfnamefont {H.~A.}\ \bibnamefont {Harrington}}, \ and\ \bibinfo {author} {\bibfnamefont {M.~T.}\ \bibnamefont {Schaub}},\ }\href@noop {} {\bibfield  {journal} {\bibinfo  {journal} {SIAM Rev.}\ }\textbf {\bibinfo {volume} {65}},\ \bibinfo {pages} {686} (\bibinfo {year} {2023})}\BibitemShut {NoStop}%
\bibitem [{\citenamefont {Boccaletti}\ \emph {et~al.}(2023)\citenamefont {Boccaletti}, \citenamefont {De~Lellis}, \citenamefont {Del~Genio}, \citenamefont {Alfaro-Bittner}, \citenamefont {Criado}, \citenamefont {Jalan},\ and\ \citenamefont {Romance}}]{boccaletti2023structure}%
  \BibitemOpen
  \bibfield  {author} {\bibinfo {author} {\bibfnamefont {S.}~\bibnamefont {Boccaletti}}, \bibinfo {author} {\bibfnamefont {P.}~\bibnamefont {De~Lellis}}, \bibinfo {author} {\bibfnamefont {C.}~\bibnamefont {Del~Genio}}, \bibinfo {author} {\bibfnamefont {K.}~\bibnamefont {Alfaro-Bittner}}, \bibinfo {author} {\bibfnamefont {R.}~\bibnamefont {Criado}}, \bibinfo {author} {\bibfnamefont {S.}~\bibnamefont {Jalan}}, \ and\ \bibinfo {author} {\bibfnamefont {M.}~\bibnamefont {Romance}},\ }\href@noop {} {\bibfield  {journal} {\bibinfo  {journal} {Phys. Rep.}\ }\textbf {\bibinfo {volume} {1018}},\ \bibinfo {pages} {1} (\bibinfo {year} {2023})}\BibitemShut {NoStop}%
\bibitem [{\citenamefont {Muolo}\ \emph {et~al.}(2024)\citenamefont {Muolo}, \citenamefont {Giambagli}, \citenamefont {Nakao}, \citenamefont {Fanelli},\ and\ \citenamefont {Carletti}}]{muolo2024turing}%
  \BibitemOpen
  \bibfield  {author} {\bibinfo {author} {\bibfnamefont {R.}~\bibnamefont {Muolo}}, \bibinfo {author} {\bibfnamefont {L.}~\bibnamefont {Giambagli}}, \bibinfo {author} {\bibfnamefont {H.}~\bibnamefont {Nakao}}, \bibinfo {author} {\bibfnamefont {D.}~\bibnamefont {Fanelli}}, \ and\ \bibinfo {author} {\bibfnamefont {T.}~\bibnamefont {Carletti}},\ }\href@noop {} {\bibfield  {journal} {\bibinfo  {journal} {Proceedings of the Royal Society A}\ }\textbf {\bibinfo {volume} {480}},\ \bibinfo {pages} {20240235} (\bibinfo {year} {2024})}\BibitemShut {NoStop}%
\bibitem [{\citenamefont {Mill{\'a}n}\ \emph {et~al.}(2025)\citenamefont {Mill{\'a}n}, \citenamefont {Sun}, \citenamefont {Giambagli}, \citenamefont {Muolo}, \citenamefont {Carletti}, \citenamefont {Torres}, \citenamefont {Radicchi}, \citenamefont {Kurths},\ and\ \citenamefont {Bianconi}}]{millan2025topology}%
  \BibitemOpen
  \bibfield  {author} {\bibinfo {author} {\bibfnamefont {A.}~\bibnamefont {Mill{\'a}n}}, \bibinfo {author} {\bibfnamefont {H.}~\bibnamefont {Sun}}, \bibinfo {author} {\bibfnamefont {L.}~\bibnamefont {Giambagli}}, \bibinfo {author} {\bibfnamefont {R.}~\bibnamefont {Muolo}}, \bibinfo {author} {\bibfnamefont {T.}~\bibnamefont {Carletti}}, \bibinfo {author} {\bibfnamefont {J.}~\bibnamefont {Torres}}, \bibinfo {author} {\bibfnamefont {F.}~\bibnamefont {Radicchi}}, \bibinfo {author} {\bibfnamefont {J.}~\bibnamefont {Kurths}}, \ and\ \bibinfo {author} {\bibfnamefont {G.}~\bibnamefont {Bianconi}},\ }\href@noop {} {\bibfield  {journal} {\bibinfo  {journal} {Nature Physics}\ }\textbf {\bibinfo {volume} {21}},\ \bibinfo {pages} {353} (\bibinfo {year} {2025})}\BibitemShut {NoStop}%
\bibitem [{\citenamefont {Battiston}\ \emph {et~al.}(2025)\citenamefont {Battiston}, \citenamefont {Bick}, \citenamefont {Lucas}, \citenamefont {Mill{\'a}n}, \citenamefont {Skardal},\ and\ \citenamefont {Zhang}}]{battiston2025collective}%
  \BibitemOpen
  \bibfield  {author} {\bibinfo {author} {\bibfnamefont {F.}~\bibnamefont {Battiston}}, \bibinfo {author} {\bibfnamefont {C.}~\bibnamefont {Bick}}, \bibinfo {author} {\bibfnamefont {M.}~\bibnamefont {Lucas}}, \bibinfo {author} {\bibfnamefont {A.}~\bibnamefont {Mill{\'a}n}}, \bibinfo {author} {\bibfnamefont {P.}~\bibnamefont {Skardal}}, \ and\ \bibinfo {author} {\bibfnamefont {Y.}~\bibnamefont {Zhang}},\ }\href@noop {} {\bibfield  {journal} {\bibinfo  {journal} {arXiv preprint arXiv:2510.05253}\ } (\bibinfo {year} {2025})}\BibitemShut {NoStop}%
\bibitem [{\citenamefont {Millán}\ \emph {et~al.}(2020)\citenamefont {Millán}, \citenamefont {Torres},\ and\ \citenamefont {Bianconi}}]{millan2020explosive}%
  \BibitemOpen
  \bibfield  {author} {\bibinfo {author} {\bibfnamefont {A.}~\bibnamefont {Millán}}, \bibinfo {author} {\bibfnamefont {J.}~\bibnamefont {Torres}}, \ and\ \bibinfo {author} {\bibfnamefont {G.}~\bibnamefont {Bianconi}},\ }\href@noop {} {\bibfield  {journal} {\bibinfo  {journal} {Phys. Rev. Lett.}\ }\textbf {\bibinfo {volume} {124}},\ \bibinfo {pages} {218301} (\bibinfo {year} {2020})}\BibitemShut {NoStop}%
\bibitem [{\citenamefont {Gambuzza}\ \emph {et~al.}(2021)\citenamefont {Gambuzza}, \citenamefont {Di~Patti}, \citenamefont {Gallo}, \citenamefont {Lepri}, \citenamefont {Romance}, \citenamefont {Criado}, \citenamefont {Frasca}, \citenamefont {Latora},\ and\ \citenamefont {Boccaletti}}]{gambuzza2021stability}%
  \BibitemOpen
  \bibfield  {author} {\bibinfo {author} {\bibfnamefont {L.}~\bibnamefont {Gambuzza}}, \bibinfo {author} {\bibfnamefont {F.}~\bibnamefont {Di~Patti}}, \bibinfo {author} {\bibfnamefont {L.}~\bibnamefont {Gallo}}, \bibinfo {author} {\bibfnamefont {S.}~\bibnamefont {Lepri}}, \bibinfo {author} {\bibfnamefont {M.}~\bibnamefont {Romance}}, \bibinfo {author} {\bibfnamefont {R.}~\bibnamefont {Criado}}, \bibinfo {author} {\bibfnamefont {M.}~\bibnamefont {Frasca}}, \bibinfo {author} {\bibfnamefont {V.}~\bibnamefont {Latora}}, \ and\ \bibinfo {author} {\bibfnamefont {S.}~\bibnamefont {Boccaletti}},\ }\href@noop {} {\bibfield  {journal} {\bibinfo  {journal} {Nat. Comm.}\ }\textbf {\bibinfo {volume} {12}},\ \bibinfo {pages} {1} (\bibinfo {year} {2021})}\BibitemShut {NoStop}%
\bibitem [{\citenamefont {Gallo}\ \emph {et~al.}(2022)\citenamefont {Gallo}, \citenamefont {Muolo}, \citenamefont {Gambuzza}, \citenamefont {Latora}, \citenamefont {Frasca},\ and\ \citenamefont {Carletti}}]{gallo2022synchronization}%
  \BibitemOpen
  \bibfield  {author} {\bibinfo {author} {\bibfnamefont {L.}~\bibnamefont {Gallo}}, \bibinfo {author} {\bibfnamefont {R.}~\bibnamefont {Muolo}}, \bibinfo {author} {\bibfnamefont {L.}~\bibnamefont {Gambuzza}}, \bibinfo {author} {\bibfnamefont {V.}~\bibnamefont {Latora}}, \bibinfo {author} {\bibfnamefont {M.}~\bibnamefont {Frasca}}, \ and\ \bibinfo {author} {\bibfnamefont {T.}~\bibnamefont {Carletti}},\ }\href@noop {} {\bibfield  {journal} {\bibinfo  {journal} {Comm. Phys.}\ }\textbf {\bibinfo {volume} {5}},\ \bibinfo {pages} {236} (\bibinfo {year} {2022})}\BibitemShut {NoStop}%
\bibitem [{\citenamefont {Von Der~Gracht}\ \emph {et~al.}(2023)\citenamefont {Von Der~Gracht}, \citenamefont {Nijholt},\ and\ \citenamefont {Rink}}]{von2023hypernetworks}%
  \BibitemOpen
  \bibfield  {author} {\bibinfo {author} {\bibfnamefont {S.}~\bibnamefont {Von Der~Gracht}}, \bibinfo {author} {\bibfnamefont {E.}~\bibnamefont {Nijholt}}, \ and\ \bibinfo {author} {\bibfnamefont {B.}~\bibnamefont {Rink}},\ }\href@noop {} {\bibfield  {journal} {\bibinfo  {journal} {SIAM Journal on Applied Mathematics}\ }\textbf {\bibinfo {volume} {83}},\ \bibinfo {pages} {2329} (\bibinfo {year} {2023})}\BibitemShut {NoStop}%
\bibitem [{\citenamefont {von~der Gracht}\ \emph {et~al.}(2024)\citenamefont {von~der Gracht}, \citenamefont {Nijholt},\ and\ \citenamefont {Rink}}]{von2024higher}%
  \BibitemOpen
  \bibfield  {author} {\bibinfo {author} {\bibfnamefont {S.}~\bibnamefont {von~der Gracht}}, \bibinfo {author} {\bibfnamefont {E.}~\bibnamefont {Nijholt}}, \ and\ \bibinfo {author} {\bibfnamefont {B.}~\bibnamefont {Rink}},\ }\href@noop {} {\bibfield  {journal} {\bibinfo  {journal} {Proceedings A of the Royal Society}\ }\textbf {\bibinfo {volume} {480}},\ \bibinfo {pages} {20230945} (\bibinfo {year} {2024})}\BibitemShut {NoStop}%
\bibitem [{\citenamefont {Zhang}\ \emph {et~al.}(2024)\citenamefont {Zhang}, \citenamefont {Skardal}, \citenamefont {Battiston}, \citenamefont {Petri},\ and\ \citenamefont {Lucas}}]{zhang2024deeper}%
  \BibitemOpen
  \bibfield  {author} {\bibinfo {author} {\bibfnamefont {Y.}~\bibnamefont {Zhang}}, \bibinfo {author} {\bibfnamefont {P.}~\bibnamefont {Skardal}}, \bibinfo {author} {\bibfnamefont {F.}~\bibnamefont {Battiston}}, \bibinfo {author} {\bibfnamefont {G.}~\bibnamefont {Petri}}, \ and\ \bibinfo {author} {\bibfnamefont {M.}~\bibnamefont {Lucas}},\ }\href@noop {} {\bibfield  {journal} {\bibinfo  {journal} {Science Advances}\ }\textbf {\bibinfo {volume} {10}},\ \bibinfo {pages} {eado8049} (\bibinfo {year} {2024})}\BibitemShut {NoStop}%
\bibitem [{\citenamefont {Anwar}\ \emph {et~al.}(2024)\citenamefont {Anwar}, \citenamefont {Sar}, \citenamefont {Perc},\ and\ \citenamefont {Ghosh}}]{anwar2024collective}%
  \BibitemOpen
  \bibfield  {author} {\bibinfo {author} {\bibfnamefont {M.~S.}\ \bibnamefont {Anwar}}, \bibinfo {author} {\bibfnamefont {G.~K.}\ \bibnamefont {Sar}}, \bibinfo {author} {\bibfnamefont {M.}~\bibnamefont {Perc}}, \ and\ \bibinfo {author} {\bibfnamefont {D.}~\bibnamefont {Ghosh}},\ }\href@noop {} {\bibfield  {journal} {\bibinfo  {journal} {Communications Physics}\ }\textbf {\bibinfo {volume} {7}},\ \bibinfo {pages} {59} (\bibinfo {year} {2024})}\BibitemShut {NoStop}%
\bibitem [{\citenamefont {Hu}\ \emph {et~al.}(2025)\citenamefont {Hu}, \citenamefont {Yu}, \citenamefont {Li}, \citenamefont {Wu}, \citenamefont {Ding},\ and\ \citenamefont {Jia}}]{hu2025effect}%
  \BibitemOpen
  \bibfield  {author} {\bibinfo {author} {\bibfnamefont {Y.}~\bibnamefont {Hu}}, \bibinfo {author} {\bibfnamefont {D.}~\bibnamefont {Yu}}, \bibinfo {author} {\bibfnamefont {T.}~\bibnamefont {Li}}, \bibinfo {author} {\bibfnamefont {Y.}~\bibnamefont {Wu}}, \bibinfo {author} {\bibfnamefont {Q.}~\bibnamefont {Ding}}, \ and\ \bibinfo {author} {\bibfnamefont {Y.}~\bibnamefont {Jia}},\ }\href@noop {} {\bibfield  {journal} {\bibinfo  {journal} {Communications Physics}\ }\textbf {\bibinfo {volume} {8}},\ \bibinfo {pages} {177} (\bibinfo {year} {2025})}\BibitemShut {NoStop}%
\bibitem [{\citenamefont {Le{\'o}n}\ \emph {et~al.}(2025)\citenamefont {Le{\'o}n}, \citenamefont {Muolo}, \citenamefont {Nakao},\ and\ \citenamefont {Taga}}]{leon2025collective}%
  \BibitemOpen
  \bibfield  {author} {\bibinfo {author} {\bibfnamefont {I.}~\bibnamefont {Le{\'o}n}}, \bibinfo {author} {\bibfnamefont {R.}~\bibnamefont {Muolo}}, \bibinfo {author} {\bibfnamefont {H.}~\bibnamefont {Nakao}}, \ and\ \bibinfo {author} {\bibfnamefont {K.}~\bibnamefont {Taga}},\ }\href@noop {} {\bibfield  {journal} {\bibinfo  {journal} {arXiv preprint arXiv:2512.19318}\ } (\bibinfo {year} {2025})}\BibitemShut {NoStop}%
\bibitem [{\citenamefont {Carletti}\ \emph {et~al.}(2020{\natexlab{a}})\citenamefont {Carletti}, \citenamefont {Battiston}, \citenamefont {Cencetti},\ and\ \citenamefont {Fanelli}}]{carletti2020random}%
  \BibitemOpen
  \bibfield  {author} {\bibinfo {author} {\bibfnamefont {T.}~\bibnamefont {Carletti}}, \bibinfo {author} {\bibfnamefont {F.}~\bibnamefont {Battiston}}, \bibinfo {author} {\bibfnamefont {G.}~\bibnamefont {Cencetti}}, \ and\ \bibinfo {author} {\bibfnamefont {D.}~\bibnamefont {Fanelli}},\ }\href@noop {} {\bibfield  {journal} {\bibinfo  {journal} {Phys. Rev. E}\ }\textbf {\bibinfo {volume} {101}},\ \bibinfo {pages} {022308} (\bibinfo {year} {2020}{\natexlab{a}})}\BibitemShut {NoStop}%
\bibitem [{\citenamefont {Schaub}\ \emph {et~al.}(2020)\citenamefont {Schaub}, \citenamefont {Benson}, \citenamefont {Horn}, \citenamefont {Lippner},\ and\ \citenamefont {Jadbabaie}}]{schaub2020random}%
  \BibitemOpen
  \bibfield  {author} {\bibinfo {author} {\bibfnamefont {M.}~\bibnamefont {Schaub}}, \bibinfo {author} {\bibfnamefont {A.}~\bibnamefont {Benson}}, \bibinfo {author} {\bibfnamefont {P.}~\bibnamefont {Horn}}, \bibinfo {author} {\bibfnamefont {G.}~\bibnamefont {Lippner}}, \ and\ \bibinfo {author} {\bibfnamefont {A.}~\bibnamefont {Jadbabaie}},\ }\href@noop {} {\bibfield  {journal} {\bibinfo  {journal} {SIAM Rev.}\ }\textbf {\bibinfo {volume} {62}},\ \bibinfo {pages} {353} (\bibinfo {year} {2020})}\BibitemShut {NoStop}%
\bibitem [{\citenamefont {Wang}\ \emph {et~al.}(2026)\citenamefont {Wang}, \citenamefont {Zhu},\ and\ \citenamefont {Liu}}]{wang2025network}%
  \BibitemOpen
  \bibfield  {author} {\bibinfo {author} {\bibfnamefont {Z.}~\bibnamefont {Wang}}, \bibinfo {author} {\bibfnamefont {J.}~\bibnamefont {Zhu}}, \ and\ \bibinfo {author} {\bibfnamefont {X.}~\bibnamefont {Liu}},\ }\href@noop {} {\bibfield  {journal} {\bibinfo  {journal} {Proceedings of the Royal Society A}\ }\textbf {\bibinfo {volume} {in press}} (\bibinfo {year} {2026})}\BibitemShut {NoStop}%
\bibitem [{\citenamefont {Muolo}\ \emph {et~al.}(2023)\citenamefont {Muolo}, \citenamefont {Gallo}, \citenamefont {Latora}, \citenamefont {Frasca},\ and\ \citenamefont {Carletti}}]{Muolo2023turing}%
  \BibitemOpen
  \bibfield  {author} {\bibinfo {author} {\bibfnamefont {R.}~\bibnamefont {Muolo}}, \bibinfo {author} {\bibfnamefont {L.}~\bibnamefont {Gallo}}, \bibinfo {author} {\bibfnamefont {V.}~\bibnamefont {Latora}}, \bibinfo {author} {\bibfnamefont {M.}~\bibnamefont {Frasca}}, \ and\ \bibinfo {author} {\bibfnamefont {T.}~\bibnamefont {Carletti}},\ }\href@noop {} {\bibfield  {journal} {\bibinfo  {journal} {Chaos, Solitons \& Fractals}\ }\textbf {\bibinfo {volume} {166}},\ \bibinfo {pages} {112912} (\bibinfo {year} {2023})}\BibitemShut {NoStop}%
\bibitem [{\citenamefont {Carletti}\ \emph {et~al.}(2020{\natexlab{b}})\citenamefont {Carletti}, \citenamefont {Fanelli},\ and\ \citenamefont {Nicoletti}}]{carletti2020dynamical}%
  \BibitemOpen
  \bibfield  {author} {\bibinfo {author} {\bibfnamefont {T.}~\bibnamefont {Carletti}}, \bibinfo {author} {\bibfnamefont {D.}~\bibnamefont {Fanelli}}, \ and\ \bibinfo {author} {\bibfnamefont {S.}~\bibnamefont {Nicoletti}},\ }\href@noop {} {\bibfield  {journal} {\bibinfo  {journal} {J. Phys. Complex.}\ }\textbf {\bibinfo {volume} {1}},\ \bibinfo {pages} {035006} (\bibinfo {year} {2020}{\natexlab{b}})}\BibitemShut {NoStop}%
\bibitem [{\citenamefont {Iacopini}\ \emph {et~al.}(2019)\citenamefont {Iacopini}, \citenamefont {Petri}, \citenamefont {Barrat},\ and\ \citenamefont {Latora}}]{iacopini2019simplicial}%
  \BibitemOpen
  \bibfield  {author} {\bibinfo {author} {\bibfnamefont {I.}~\bibnamefont {Iacopini}}, \bibinfo {author} {\bibfnamefont {G.}~\bibnamefont {Petri}}, \bibinfo {author} {\bibfnamefont {A.}~\bibnamefont {Barrat}}, \ and\ \bibinfo {author} {\bibfnamefont {V.}~\bibnamefont {Latora}},\ }\href@noop {} {\bibfield  {journal} {\bibinfo  {journal} {Nat. Comm.}\ }\textbf {\bibinfo {volume} {10}},\ \bibinfo {pages} {2485} (\bibinfo {year} {2019})}\BibitemShut {NoStop}%
\bibitem [{\citenamefont {Neuh\"auser}\ \emph {et~al.}(2020)\citenamefont {Neuh\"auser}, \citenamefont {Mellor},\ and\ \citenamefont {Lambiotte}}]{Neuhauser2020multibody}%
  \BibitemOpen
  \bibfield  {author} {\bibinfo {author} {\bibfnamefont {L.}~\bibnamefont {Neuh\"auser}}, \bibinfo {author} {\bibfnamefont {A.}~\bibnamefont {Mellor}}, \ and\ \bibinfo {author} {\bibfnamefont {R.}~\bibnamefont {Lambiotte}},\ }\href@noop {} {\bibfield  {journal} {\bibinfo  {journal} {Phys. Rev. E}\ }\textbf {\bibinfo {volume} {101}},\ \bibinfo {pages} {032310} (\bibinfo {year} {2020})}\BibitemShut {NoStop}%
\bibitem [{\citenamefont {De~Lellis}\ \emph {et~al.}(2022)\citenamefont {De~Lellis}, \citenamefont {Della~Rossa}, \citenamefont {Iudice},\ and\ \citenamefont {Liuzza}}]{de2022pinning}%
  \BibitemOpen
  \bibfield  {author} {\bibinfo {author} {\bibfnamefont {P.}~\bibnamefont {De~Lellis}}, \bibinfo {author} {\bibfnamefont {F.}~\bibnamefont {Della~Rossa}}, \bibinfo {author} {\bibfnamefont {F.~L.}\ \bibnamefont {Iudice}}, \ and\ \bibinfo {author} {\bibfnamefont {D.}~\bibnamefont {Liuzza}},\ }\href@noop {} {\bibfield  {journal} {\bibinfo  {journal} {IEEE Control Syst. Lett.}\ }\textbf {\bibinfo {volume} {7}},\ \bibinfo {pages} {691} (\bibinfo {year} {2022})}\BibitemShut {NoStop}%
\bibitem [{\citenamefont {Della~Rossa}\ \emph {et~al.}(2023)\citenamefont {Della~Rossa}, \citenamefont {Liuzza}, \citenamefont {Lo~Iudice},\ and\ \citenamefont {De~Lellis}}]{della2023emergence}%
  \BibitemOpen
  \bibfield  {author} {\bibinfo {author} {\bibfnamefont {F.}~\bibnamefont {Della~Rossa}}, \bibinfo {author} {\bibfnamefont {D.}~\bibnamefont {Liuzza}}, \bibinfo {author} {\bibfnamefont {F.}~\bibnamefont {Lo~Iudice}}, \ and\ \bibinfo {author} {\bibfnamefont {P.}~\bibnamefont {De~Lellis}},\ }\href@noop {} {\bibfield  {journal} {\bibinfo  {journal} {Phys. Rev. Lett.}\ }\textbf {\bibinfo {volume} {131}},\ \bibinfo {pages} {207401} (\bibinfo {year} {2023})}\BibitemShut {NoStop}%
\bibitem [{\citenamefont {Xia}\ and\ \citenamefont {Xiang}(2024)}]{xia2024pinning}%
  \BibitemOpen
  \bibfield  {author} {\bibinfo {author} {\bibfnamefont {R.}~\bibnamefont {Xia}}\ and\ \bibinfo {author} {\bibfnamefont {L.}~\bibnamefont {Xiang}},\ }\href@noop {} {\bibfield  {journal} {\bibinfo  {journal} {European Journal of Control}\ }\textbf {\bibinfo {volume} {77}},\ \bibinfo {pages} {100994} (\bibinfo {year} {2024})}\BibitemShut {NoStop}%
\bibitem [{\citenamefont {Kuramoto}(1975)}]{Kuramoto1975self}%
  \BibitemOpen
  \bibfield  {author} {\bibinfo {author} {\bibfnamefont {Y.}~\bibnamefont {Kuramoto}},\ }in\ \href@noop {} {\emph {\bibinfo {booktitle} {International Symposium on Mathematical Problems in Theoretical Physics}}},\ \bibinfo {editor} {edited by\ \bibinfo {editor} {\bibfnamefont {H.}~\bibnamefont {Araki}}}\ (\bibinfo  {publisher} {Springer Berlin Heidelberg},\ \bibinfo {address} {Berlin, Heidelberg},\ \bibinfo {year} {1975})\ pp.\ \bibinfo {pages} {420--422}\BibitemShut {NoStop}%
\bibitem [{\citenamefont {Kuramoto}(1984)}]{Kuramoto1984chemical}%
  \BibitemOpen
  \bibfield  {author} {\bibinfo {author} {\bibfnamefont {Y.}~\bibnamefont {Kuramoto}},\ }\href@noop {} {\emph {\bibinfo {title} {Chemical Oscillations, Waves, and Turbulence}}}\ (\bibinfo  {publisher} {Springer},\ \bibinfo {address} {Berlin},\ \bibinfo {year} {1984})\BibitemShut {NoStop}%
\bibitem [{\citenamefont {Acebr{\'o}n}\ \emph {et~al.}(2005)\citenamefont {Acebr{\'o}n}, \citenamefont {Bonilla}, \citenamefont {P{\'e}rez~Vicente}, \citenamefont {Ritort},\ and\ \citenamefont {Spigler}}]{acebron2005kuramoto}%
  \BibitemOpen
  \bibfield  {author} {\bibinfo {author} {\bibfnamefont {J.}~\bibnamefont {Acebr{\'o}n}}, \bibinfo {author} {\bibfnamefont {L.}~\bibnamefont {Bonilla}}, \bibinfo {author} {\bibfnamefont {C.}~\bibnamefont {P{\'e}rez~Vicente}}, \bibinfo {author} {\bibfnamefont {F.}~\bibnamefont {Ritort}}, \ and\ \bibinfo {author} {\bibfnamefont {R.}~\bibnamefont {Spigler}},\ }\href@noop {} {\bibfield  {journal} {\bibinfo  {journal} {Reviews of modern physics}\ }\textbf {\bibinfo {volume} {77}},\ \bibinfo {pages} {137} (\bibinfo {year} {2005})}\BibitemShut {NoStop}%
\bibitem [{\citenamefont {Tanaka}\ and\ \citenamefont {Aoyagi}(2011)}]{tanaka2011multistable}%
  \BibitemOpen
  \bibfield  {author} {\bibinfo {author} {\bibfnamefont {T.}~\bibnamefont {Tanaka}}\ and\ \bibinfo {author} {\bibfnamefont {T.}~\bibnamefont {Aoyagi}},\ }\href {\doibase 10.1103/PhysRevLett.106.224101} {\bibfield  {journal} {\bibinfo  {journal} {Phys. Rev. Lett.}\ }\textbf {\bibinfo {volume} {106}},\ \bibinfo {pages} {224101} (\bibinfo {year} {2011})}\BibitemShut {NoStop}%
\bibitem [{\citenamefont {Skardal}\ and\ \citenamefont {Arenas}(2020)}]{Skardal2020higher}%
  \BibitemOpen
  \bibfield  {author} {\bibinfo {author} {\bibfnamefont {P.}~\bibnamefont {Skardal}}\ and\ \bibinfo {author} {\bibfnamefont {A.}~\bibnamefont {Arenas}},\ }\href@noop {} {\bibfield  {journal} {\bibinfo  {journal} {Communications Physics}\ }\textbf {\bibinfo {volume} {3}},\ \bibinfo {pages} {218} (\bibinfo {year} {2020})}\BibitemShut {NoStop}%
\bibitem [{\citenamefont {Lucas}\ \emph {et~al.}(2020)\citenamefont {Lucas}, \citenamefont {Cencetti},\ and\ \citenamefont {Battiston}}]{lucas2020multiorder}%
  \BibitemOpen
  \bibfield  {author} {\bibinfo {author} {\bibfnamefont {M.}~\bibnamefont {Lucas}}, \bibinfo {author} {\bibfnamefont {G.}~\bibnamefont {Cencetti}}, \ and\ \bibinfo {author} {\bibfnamefont {F.}~\bibnamefont {Battiston}},\ }\href@noop {} {\bibfield  {journal} {\bibinfo  {journal} {Physical Review Research}\ }\textbf {\bibinfo {volume} {2}},\ \bibinfo {pages} {033410} (\bibinfo {year} {2020})}\BibitemShut {NoStop}%
\bibitem [{\citenamefont {Le\'{o}n}\ \emph {et~al.}(2024)\citenamefont {Le\'{o}n}, \citenamefont {Muolo}, \citenamefont {Hata},\ and\ \citenamefont {Nakao}}]{Leon2024higher}%
  \BibitemOpen
  \bibfield  {author} {\bibinfo {author} {\bibfnamefont {I.}~\bibnamefont {Le\'{o}n}}, \bibinfo {author} {\bibfnamefont {R.}~\bibnamefont {Muolo}}, \bibinfo {author} {\bibfnamefont {S.}~\bibnamefont {Hata}}, \ and\ \bibinfo {author} {\bibfnamefont {H.}~\bibnamefont {Nakao}},\ }\href@noop {} {\bibfield  {journal} {\bibinfo  {journal} {Chaos: An Interdisciplinary Journal of Nonlinear Science}\ }\textbf {\bibinfo {volume} {34}},\ \bibinfo {pages} {013105} (\bibinfo {year} {2024})}\BibitemShut {NoStop}%
\bibitem [{\citenamefont {Le\'{o}n}\ \emph {et~al.}(2025)\citenamefont {Le\'{o}n}, \citenamefont {Muolo}, \citenamefont {Hata},\ and\ \citenamefont {Nakao}}]{Leon2025theory}%
  \BibitemOpen
  \bibfield  {author} {\bibinfo {author} {\bibfnamefont {I.}~\bibnamefont {Le\'{o}n}}, \bibinfo {author} {\bibfnamefont {R.}~\bibnamefont {Muolo}}, \bibinfo {author} {\bibfnamefont {S.}~\bibnamefont {Hata}}, \ and\ \bibinfo {author} {\bibfnamefont {H.}~\bibnamefont {Nakao}},\ }\href@noop {} {\bibfield  {journal} {\bibinfo  {journal} {Physica D: Nonlinear Phenomena}\ }\textbf {\bibinfo {volume} {482}},\ \bibinfo {pages} {134858} (\bibinfo {year} {2025})}\BibitemShut {NoStop}%
\bibitem [{\citenamefont {Nakao}(2016)}]{Nakao2016phase}%
  \BibitemOpen
  \bibfield  {author} {\bibinfo {author} {\bibfnamefont {H.}~\bibnamefont {Nakao}},\ }\href {\doibase 10.1080/00107514.2015.1094987} {\bibfield  {journal} {\bibinfo  {journal} {Contemporary Physics}\ }\textbf {\bibinfo {volume} {57}},\ \bibinfo {pages} {188} (\bibinfo {year} {2016})}\BibitemShut {NoStop}%
\bibitem [{\citenamefont {Monga}\ \emph {et~al.}(2019)\citenamefont {Monga}, \citenamefont {Wilson}, \citenamefont {Matchen},\ and\ \citenamefont {Moehlis}}]{Monga2019phase1}%
  \BibitemOpen
  \bibfield  {author} {\bibinfo {author} {\bibfnamefont {B.}~\bibnamefont {Monga}}, \bibinfo {author} {\bibfnamefont {D.}~\bibnamefont {Wilson}}, \bibinfo {author} {\bibfnamefont {T.}~\bibnamefont {Matchen}}, \ and\ \bibinfo {author} {\bibfnamefont {J.}~\bibnamefont {Moehlis}},\ }\href@noop {} {\bibfield  {journal} {\bibinfo  {journal} {Biological cybernetics}\ }\textbf {\bibinfo {volume} {113}},\ \bibinfo {pages} {11} (\bibinfo {year} {2019})}\BibitemShut {NoStop}%
\bibitem [{\citenamefont {Pietras}\ and\ \citenamefont {Daffertshofer}(2019)}]{pietras2019network}%
  \BibitemOpen
  \bibfield  {author} {\bibinfo {author} {\bibfnamefont {B.}~\bibnamefont {Pietras}}\ and\ \bibinfo {author} {\bibfnamefont {A.}~\bibnamefont {Daffertshofer}},\ }\href@noop {} {\bibfield  {journal} {\bibinfo  {journal} {Physics Reports}\ }\textbf {\bibinfo {volume} {819}},\ \bibinfo {pages} {1} (\bibinfo {year} {2019})}\BibitemShut {NoStop}%
\bibitem [{\citenamefont {Kuramoto}\ and\ \citenamefont {Nakao}(2019)}]{Kuramoto_2019}%
  \BibitemOpen
  \bibfield  {author} {\bibinfo {author} {\bibfnamefont {Y.}~\bibnamefont {Kuramoto}}\ and\ \bibinfo {author} {\bibfnamefont {H.}~\bibnamefont {Nakao}},\ }\href {\doibase 10.1098/rsta.2019.0041} {\bibfield  {journal} {\bibinfo  {journal} {Philosophical Transactions of the Royal Society A: Mathematical, Physical and Engineering Sciences}\ }\textbf {\bibinfo {volume} {377}},\ \bibinfo {pages} {20190041} (\bibinfo {year} {2019})}\BibitemShut {NoStop}%
\bibitem [{\citenamefont {Le\'on}\ and\ \citenamefont {Paz\'o}(2019)}]{Leon2019phase}%
  \BibitemOpen
  \bibfield  {author} {\bibinfo {author} {\bibfnamefont {I.}~\bibnamefont {Le\'on}}\ and\ \bibinfo {author} {\bibfnamefont {D.}~\bibnamefont {Paz\'o}},\ }\href@noop {} {\bibfield  {journal} {\bibinfo  {journal} {Phys. Rev. E}\ }\textbf {\bibinfo {volume} {100}},\ \bibinfo {pages} {012211} (\bibinfo {year} {2019})}\BibitemShut {NoStop}%
\bibitem [{\citenamefont {Gengel}\ \emph {et~al.}(2020)\citenamefont {Gengel}, \citenamefont {Teichmann}, \citenamefont {Rosenblum},\ and\ \citenamefont {Pikovsky}}]{gengel2020high}%
  \BibitemOpen
  \bibfield  {author} {\bibinfo {author} {\bibfnamefont {E.}~\bibnamefont {Gengel}}, \bibinfo {author} {\bibfnamefont {E.}~\bibnamefont {Teichmann}}, \bibinfo {author} {\bibfnamefont {M.}~\bibnamefont {Rosenblum}}, \ and\ \bibinfo {author} {\bibfnamefont {A.}~\bibnamefont {Pikovsky}},\ }\href@noop {} {\bibfield  {journal} {\bibinfo  {journal} {Journal of Physics: Complexity}\ }\textbf {\bibinfo {volume} {2}},\ \bibinfo {pages} {015005} (\bibinfo {year} {2020})}\BibitemShut {NoStop}%
\bibitem [{\citenamefont {Nijholt}\ \emph {et~al.}(2022)\citenamefont {Nijholt}, \citenamefont {Ocampo-Espindola}, \citenamefont {Eroglu}, \citenamefont {Kiss},\ and\ \citenamefont {Pereira}}]{nijholt2022emergent}%
  \BibitemOpen
  \bibfield  {author} {\bibinfo {author} {\bibfnamefont {E.}~\bibnamefont {Nijholt}}, \bibinfo {author} {\bibfnamefont {J.~L.}\ \bibnamefont {Ocampo-Espindola}}, \bibinfo {author} {\bibfnamefont {D.}~\bibnamefont {Eroglu}}, \bibinfo {author} {\bibfnamefont {I.~Z.}\ \bibnamefont {Kiss}}, \ and\ \bibinfo {author} {\bibfnamefont {T.}~\bibnamefont {Pereira}},\ }\href@noop {} {\bibfield  {journal} {\bibinfo  {journal} {Nature communications}\ }\textbf {\bibinfo {volume} {13}},\ \bibinfo {pages} {4849} (\bibinfo {year} {2022})}\BibitemShut {NoStop}%
\bibitem [{\citenamefont {Mau}\ \emph {et~al.}(2023)\citenamefont {Mau}, \citenamefont {Rosenblum},\ and\ \citenamefont {Pikovsky}}]{mau2023high}%
  \BibitemOpen
  \bibfield  {author} {\bibinfo {author} {\bibfnamefont {E.~T.}\ \bibnamefont {Mau}}, \bibinfo {author} {\bibfnamefont {M.}~\bibnamefont {Rosenblum}}, \ and\ \bibinfo {author} {\bibfnamefont {A.}~\bibnamefont {Pikovsky}},\ }\href@noop {} {\bibfield  {journal} {\bibinfo  {journal} {Chaos: an interdisciplinary journal of nonlinear science}\ }\textbf {\bibinfo {volume} {33}} (\bibinfo {year} {2023})}\BibitemShut {NoStop}%
\bibitem [{\citenamefont {Mau}\ \emph {et~al.}(2024)\citenamefont {Mau}, \citenamefont {Omel'chenko},\ and\ \citenamefont {Rosenblum}}]{mau2024phase}%
  \BibitemOpen
  \bibfield  {author} {\bibinfo {author} {\bibfnamefont {E.~T.}\ \bibnamefont {Mau}}, \bibinfo {author} {\bibfnamefont {O.~E.}\ \bibnamefont {Omel'chenko}}, \ and\ \bibinfo {author} {\bibfnamefont {M.}~\bibnamefont {Rosenblum}},\ }\href@noop {} {\bibfield  {journal} {\bibinfo  {journal} {Physical Review E}\ }\textbf {\bibinfo {volume} {110}},\ \bibinfo {pages} {L022201} (\bibinfo {year} {2024})}\BibitemShut {NoStop}%
\bibitem [{\citenamefont {Bick}\ \emph {et~al.}(2024{\natexlab{a}})\citenamefont {Bick}, \citenamefont {B{\"o}hle},\ and\ \citenamefont {Kuehn}}]{bick2024higher}%
  \BibitemOpen
  \bibfield  {author} {\bibinfo {author} {\bibfnamefont {C.}~\bibnamefont {Bick}}, \bibinfo {author} {\bibfnamefont {T.}~\bibnamefont {B{\"o}hle}}, \ and\ \bibinfo {author} {\bibfnamefont {C.}~\bibnamefont {Kuehn}},\ }\href@noop {} {\bibfield  {journal} {\bibinfo  {journal} {Journal of Nonlinear Science}\ }\textbf {\bibinfo {volume} {34}},\ \bibinfo {pages} {77} (\bibinfo {year} {2024}{\natexlab{a}})}\BibitemShut {NoStop}%
\bibitem [{\citenamefont {von~der Gracht}\ \emph {et~al.}(2023)\citenamefont {von~der Gracht}, \citenamefont {Nijholt},\ and\ \citenamefont {Rink}}]{von2023parametrisation}%
  \BibitemOpen
  \bibfield  {author} {\bibinfo {author} {\bibfnamefont {S.}~\bibnamefont {von~der Gracht}}, \bibinfo {author} {\bibfnamefont {E.}~\bibnamefont {Nijholt}}, \ and\ \bibinfo {author} {\bibfnamefont {B.}~\bibnamefont {Rink}},\ }\href@noop {} {\bibfield  {journal} {\bibinfo  {journal} {arXiv preprint arXiv:2306.03320}\ } (\bibinfo {year} {2023})}\BibitemShut {NoStop}%
\bibitem [{\citenamefont {Bick}\ \emph {et~al.}(2024{\natexlab{b}})\citenamefont {Bick}, \citenamefont {Rink},\ and\ \citenamefont {de~Wolff}}]{bick2024time}%
  \BibitemOpen
  \bibfield  {author} {\bibinfo {author} {\bibfnamefont {C.}~\bibnamefont {Bick}}, \bibinfo {author} {\bibfnamefont {B.}~\bibnamefont {Rink}}, \ and\ \bibinfo {author} {\bibfnamefont {B.}~\bibnamefont {de~Wolff}},\ }\href@noop {} {\bibfield  {journal} {\bibinfo  {journal} {arXiv preprint arXiv:2404.11340}\ } (\bibinfo {year} {2024}{\natexlab{b}})}\BibitemShut {NoStop}%
\bibitem [{\citenamefont {Bick}\ \emph {et~al.}(2025)\citenamefont {Bick}, \citenamefont {Rink},\ and\ \citenamefont {de~Wolff}}]{bick2025higher}%
  \BibitemOpen
  \bibfield  {author} {\bibinfo {author} {\bibfnamefont {C.}~\bibnamefont {Bick}}, \bibinfo {author} {\bibfnamefont {B.}~\bibnamefont {Rink}}, \ and\ \bibinfo {author} {\bibfnamefont {B.}~\bibnamefont {de~Wolff}},\ }\href@noop {} {\bibfield  {journal} {\bibinfo  {journal} {arXiv preprint arXiv:2510.27524}\ } (\bibinfo {year} {2025})}\BibitemShut {NoStop}%
\bibitem [{\citenamefont {Yeung}\ and\ \citenamefont {Strogatz}(1999)}]{Yeung_1999}%
  \BibitemOpen
  \bibfield  {author} {\bibinfo {author} {\bibfnamefont {M.~K.~S.}\ \bibnamefont {Yeung}}\ and\ \bibinfo {author} {\bibfnamefont {S.~H.}\ \bibnamefont {Strogatz}},\ }\href {\doibase 10.1103/PhysRevLett.82.648} {\bibfield  {journal} {\bibinfo  {journal} {Phys. Rev. Lett.}\ }\textbf {\bibinfo {volume} {82}},\ \bibinfo {pages} {648} (\bibinfo {year} {1999})}\BibitemShut {NoStop}%
\bibitem [{\citenamefont {Ko}\ and\ \citenamefont {Ermentrout}(2007)}]{ermentrout2007delay}%
  \BibitemOpen
  \bibfield  {author} {\bibinfo {author} {\bibfnamefont {T.-W.}\ \bibnamefont {Ko}}\ and\ \bibinfo {author} {\bibfnamefont {G.~B.}\ \bibnamefont {Ermentrout}},\ }\href {\doibase 10.1103/PhysRevE.76.056206} {\bibfield  {journal} {\bibinfo  {journal} {Phys. Rev. E}\ }\textbf {\bibinfo {volume} {76}},\ \bibinfo {pages} {056206} (\bibinfo {year} {2007})}\BibitemShut {NoStop}%
\bibitem [{\citenamefont {Kozyreff}\ \emph {et~al.}(2000)\citenamefont {Kozyreff}, \citenamefont {Vladimirov},\ and\ \citenamefont {Mandel}}]{Kozyreff2000Lasers}%
  \BibitemOpen
  \bibfield  {author} {\bibinfo {author} {\bibfnamefont {G.}~\bibnamefont {Kozyreff}}, \bibinfo {author} {\bibfnamefont {A.~G.}\ \bibnamefont {Vladimirov}}, \ and\ \bibinfo {author} {\bibfnamefont {P.}~\bibnamefont {Mandel}},\ }\href {\doibase 10.1103/PhysRevLett.85.3809} {\bibfield  {journal} {\bibinfo  {journal} {Phys. Rev. Lett.}\ }\textbf {\bibinfo {volume} {85}},\ \bibinfo {pages} {3809} (\bibinfo {year} {2000})}\BibitemShut {NoStop}%
\bibitem [{\citenamefont {Takamatsu}\ \emph {et~al.}(2000)\citenamefont {Takamatsu}, \citenamefont {Fujii},\ and\ \citenamefont {Endo}}]{Takamatsu2000Living}%
  \BibitemOpen
  \bibfield  {author} {\bibinfo {author} {\bibfnamefont {A.}~\bibnamefont {Takamatsu}}, \bibinfo {author} {\bibfnamefont {T.}~\bibnamefont {Fujii}}, \ and\ \bibinfo {author} {\bibfnamefont {I.}~\bibnamefont {Endo}},\ }\href {\doibase 10.1103/PhysRevLett.85.2026} {\bibfield  {journal} {\bibinfo  {journal} {Phys. Rev. Lett.}\ }\textbf {\bibinfo {volume} {85}},\ \bibinfo {pages} {2026} (\bibinfo {year} {2000})}\BibitemShut {NoStop}%
\bibitem [{\citenamefont {Taher}\ \emph {et~al.}(2019)\citenamefont {Taher}, \citenamefont {Olmi},\ and\ \citenamefont {Sch\"oll}}]{Taher2019grid}%
  \BibitemOpen
  \bibfield  {author} {\bibinfo {author} {\bibfnamefont {H.}~\bibnamefont {Taher}}, \bibinfo {author} {\bibfnamefont {S.}~\bibnamefont {Olmi}}, \ and\ \bibinfo {author} {\bibfnamefont {E.}~\bibnamefont {Sch\"oll}},\ }\href {\doibase 10.1103/PhysRevE.100.062306} {\bibfield  {journal} {\bibinfo  {journal} {Phys. Rev. E}\ }\textbf {\bibinfo {volume} {100}},\ \bibinfo {pages} {062306} (\bibinfo {year} {2019})}\BibitemShut {NoStop}%
\bibitem [{\citenamefont {B\"{o}ttcher}\ \emph {et~al.}(2020)\citenamefont {B\"{o}ttcher}, \citenamefont {Otto}, \citenamefont {Kettemann},\ and\ \citenamefont {Agert}}]{Bottcher2020grid}%
  \BibitemOpen
  \bibfield  {author} {\bibinfo {author} {\bibfnamefont {P.~C.}\ \bibnamefont {B\"{o}ttcher}}, \bibinfo {author} {\bibfnamefont {A.}~\bibnamefont {Otto}}, \bibinfo {author} {\bibfnamefont {S.}~\bibnamefont {Kettemann}}, \ and\ \bibinfo {author} {\bibfnamefont {C.}~\bibnamefont {Agert}},\ }\href {\doibase 10.1063/1.5122738} {\bibfield  {journal} {\bibinfo  {journal} {Chaos: An Interdisciplinary Journal of Nonlinear Science}\ }\textbf {\bibinfo {volume} {30}},\ \bibinfo {pages} {013122} (\bibinfo {year} {2020})}\BibitemShut {NoStop}%
\bibitem [{\citenamefont {Ott}\ and\ \citenamefont {Antonsen}(2008)}]{Ott_Antonsen_2008}%
  \BibitemOpen
  \bibfield  {author} {\bibinfo {author} {\bibfnamefont {E.}~\bibnamefont {Ott}}\ and\ \bibinfo {author} {\bibfnamefont {T.}~\bibnamefont {Antonsen}},\ }\href {\doibase 10.1063/1.2930766} {\bibfield  {journal} {\bibinfo  {journal} {Chaos: An Interdisciplinary Journal of Nonlinear Science}\ }\textbf {\bibinfo {volume} {18}} (\bibinfo {year} {2008}),\ 10.1063/1.2930766}\BibitemShut {NoStop}%
\bibitem [{\citenamefont {Sakaguchi}\ and\ \citenamefont {Kuramoto}(1986)}]{Sakaguchi_1986}%
  \BibitemOpen
  \bibfield  {author} {\bibinfo {author} {\bibfnamefont {H.}~\bibnamefont {Sakaguchi}}\ and\ \bibinfo {author} {\bibfnamefont {Y.}~\bibnamefont {Kuramoto}},\ }\href {\doibase 10.1143/PTP.76.576} {\bibfield  {journal} {\bibinfo  {journal} {Progress of Theoretical Physics}\ }\textbf {\bibinfo {volume} {76}},\ \bibinfo {pages} {576} (\bibinfo {year} {1986})},\ \Eprint {http://arxiv.org/abs/https://academic.oup.com/ptp/article-pdf/76/3/576/5302137/76-3-576.pdf} {https://academic.oup.com/ptp/article-pdf/76/3/576/5302137/76-3-576.pdf} \BibitemShut {NoStop}%
\bibitem [{\citenamefont {Crook}\ \emph {et~al.}(1997)\citenamefont {Crook}, \citenamefont {Ermentrout}, \citenamefont {Vanier},\ and\ \citenamefont {Bower}}]{ermentrout1997delay}%
  \BibitemOpen
  \bibfield  {author} {\bibinfo {author} {\bibfnamefont {S.~M.}\ \bibnamefont {Crook}}, \bibinfo {author} {\bibfnamefont {G.~B.}\ \bibnamefont {Ermentrout}}, \bibinfo {author} {\bibfnamefont {M.~C.}\ \bibnamefont {Vanier}}, \ and\ \bibinfo {author} {\bibfnamefont {J.~M.}\ \bibnamefont {Bower}},\ }\href {\doibase 10.1023/A:1008843412952} {\bibfield  {journal} {\bibinfo  {journal} {Journal of Computational Neuroscience}\ }\textbf {\bibinfo {volume} {4}},\ \bibinfo {pages} {161} (\bibinfo {year} {1997})}\BibitemShut {NoStop}%
\bibitem [{\citenamefont {Namura}\ \emph {et~al.}(2026)\citenamefont {Namura}, \citenamefont {Muolo},\ and\ \citenamefont {Nakao}}]{namura2025optimal}%
  \BibitemOpen
  \bibfield  {author} {\bibinfo {author} {\bibfnamefont {N.}~\bibnamefont {Namura}}, \bibinfo {author} {\bibfnamefont {R.}~\bibnamefont {Muolo}}, \ and\ \bibinfo {author} {\bibfnamefont {H.}~\bibnamefont {Nakao}},\ }\href@noop {} {\bibfield  {journal} {\bibinfo  {journal} {Chaos}\ }\textbf {\bibinfo {volume} {36}},\ \bibinfo {pages} {023120} (\bibinfo {year} {2026})}\BibitemShut {NoStop}%
\bibitem [{\citenamefont {Kori}(2003)}]{Kori_2003}%
  \BibitemOpen
  \bibfield  {author} {\bibinfo {author} {\bibfnamefont {H.}~\bibnamefont {Kori}},\ }\href {\doibase 10.1103/PhysRevE.68.021919} {\bibfield  {journal} {\bibinfo  {journal} {Phys. Rev. E}\ }\textbf {\bibinfo {volume} {68}},\ \bibinfo {pages} {021919} (\bibinfo {year} {2003})}\BibitemShut {NoStop}%
\bibitem [{\citenamefont {Lee}\ \emph {et~al.}(2009)\citenamefont {Lee}, \citenamefont {Ott},\ and\ \citenamefont {Antonsen}}]{Ott_Antonsen_2009}%
  \BibitemOpen
  \bibfield  {author} {\bibinfo {author} {\bibfnamefont {W.~S.}\ \bibnamefont {Lee}}, \bibinfo {author} {\bibfnamefont {E.}~\bibnamefont {Ott}}, \ and\ \bibinfo {author} {\bibfnamefont {T.~M.}\ \bibnamefont {Antonsen}},\ }\href {\doibase 10.1103/PhysRevLett.103.044101} {\bibfield  {journal} {\bibinfo  {journal} {Phys. Rev. Lett.}\ }\textbf {\bibinfo {volume} {103}},\ \bibinfo {pages} {044101} (\bibinfo {year} {2009})}\BibitemShut {NoStop}%
\bibitem [{\citenamefont {Pikovsky}\ and\ \citenamefont {Rosenblum}(2011)}]{pikovsky2011wsansatz}%
  \BibitemOpen
  \bibfield  {author} {\bibinfo {author} {\bibfnamefont {A.}~\bibnamefont {Pikovsky}}\ and\ \bibinfo {author} {\bibfnamefont {M.}~\bibnamefont {Rosenblum}},\ }\href {\doibase https://doi.org/10.1016/j.physd.2011.01.002} {\bibfield  {journal} {\bibinfo  {journal} {Physica D: Nonlinear Phenomena}\ }\textbf {\bibinfo {volume} {240}},\ \bibinfo {pages} {872} (\bibinfo {year} {2011})}\BibitemShut {NoStop}%
\bibitem [{\citenamefont {Choi}\ \emph {et~al.}(2000)\citenamefont {Choi}, \citenamefont {Kim}, \citenamefont {Kim},\ and\ \citenamefont {Hong}}]{choi2000delay}%
  \BibitemOpen
  \bibfield  {author} {\bibinfo {author} {\bibfnamefont {M.~Y.}\ \bibnamefont {Choi}}, \bibinfo {author} {\bibfnamefont {H.~J.}\ \bibnamefont {Kim}}, \bibinfo {author} {\bibfnamefont {D.}~\bibnamefont {Kim}}, \ and\ \bibinfo {author} {\bibfnamefont {H.}~\bibnamefont {Hong}},\ }\href {\doibase 10.1103/PhysRevE.61.371} {\bibfield  {journal} {\bibinfo  {journal} {Phys. Rev. E}\ }\textbf {\bibinfo {volume} {61}},\ \bibinfo {pages} {371} (\bibinfo {year} {2000})}\BibitemShut {NoStop}%
\bibitem [{\citenamefont {Earl}\ and\ \citenamefont {Strogatz}(2003)}]{strogatz2003delay}%
  \BibitemOpen
  \bibfield  {author} {\bibinfo {author} {\bibfnamefont {M.~G.}\ \bibnamefont {Earl}}\ and\ \bibinfo {author} {\bibfnamefont {S.~H.}\ \bibnamefont {Strogatz}},\ }\href {\doibase 10.1103/PhysRevE.67.036204} {\bibfield  {journal} {\bibinfo  {journal} {Phys. Rev. E}\ }\textbf {\bibinfo {volume} {67}},\ \bibinfo {pages} {036204} (\bibinfo {year} {2003})}\BibitemShut {NoStop}%
\end{thebibliography}
\end{document}